\documentclass[reprint,amsmath,amssymb,aps, prd]{revtex4-2}
\usepackage[colorlinks=true,linkcolor=blue,citecolor=blue,urlcolor=blue]{hyperref}
\usepackage{newtxtext,newtxmath}
\usepackage[T1]{fontenc}
\usepackage{ae,aecompl}
\usepackage[dvipsnames]{xcolor}
\usepackage[left]{lineno}
\usepackage{caption}
\usepackage{subcaption}
\usepackage{graphicx}	
\usepackage{amsmath}	
\usepackage[justification=raggedright,singlelinecheck=false]{caption}
\usepackage{tabularray}
\usepackage{booktabs}
\usepackage{hyperref}

\newcommand{\jcap}{J. Cosmol. Astropart. Phys.}
\newcommand{\mnras}{Mon.\ Not.\ R.\ Astron.\ Soc.}
\newcommand{\apjl}{Astrophys.\ J.\ Letters}

\newcommand{\aap}{Astron. Astrophys.}
\newcommand{\aj}{Astron. J.}

\newcommand{\physrep}{Phys. Rep.}

\newcommand{\pasp}{Publ. Astron. Soc. Pac.}
\newcommand{\pnas}{Proc. Natl. Acad. Sci.}

\def\bfx{{\bf x}}
\def\bfk{{\bf k}}

\def\calG{{\mathcal G}}

\def\calF{{\mathcal F}}

\def\calQ{{\mathcal Q}}
\def\calL{{\mathcal L}}
\def\calP{{\mathcal P}}

\def\calO{{\mathcal O}}

\renewcommand{\vec}[1]{\boldsymbol{#1}}

\def\bfk{{\bf k}}

\def\bfr{{\bf r}}
\def\hr{{\hat{r}}}
\def\bfR{{\bf R}}

\def\calB{{\cal B}}
\def\calC{{\cal C}}
\def\calD{{\cal D}}
\def\calE{{\cal E}}
\def\calL{{\cal L}}

\def\calG{{\cal G}}

\def\calN{{\cal N}}
\def\calO{{\cal O}}
\def\calQ{{\cal Q}}
\def\calP{{\cal P}}
\def\calR{{\cal R}}

\def\bfx{{\bf x}}

\newcommand{\bfs}{\mathbf s}

\newcommand{\hn}{\hat{\vec n}}

\newcommand{\av}[1]{\left\langle{#1}\right\rangle} 

\newcommand{\jhc}[1]{\textcolor{black}{#1}} 

\def\ezmock{{\textsc{EZmock}}}
\def\abacus{{\textsc{Abacus}}}
\def\gpch{{h^{-1}\rm{Gpc}}}
\def\mpch{{h^{-1}\rm{Mpc}}}
\def\hmpc{{{\rm Mpc}^{-1} h}}

\newcommand{\six}[6]{\left(\begin{array}{ccc}
									{#1}& {#2}& {#3}\\
									{#4}& {#5}& {#6} \\
\end{array}\right)}

\newcommand{\nine}[9]{\left\{\begin{array}{ccc}
									{#1}& {#2}& {#3}\\
									{#4}& {#5}& {#6} \\
							    	{#7}& {#8}& {#9} \\
\end{array}\right\}}

\begin{document}
\label{firstpage}

\title{Study of the Connected Four-Point Correlation Function of Galaxies from DESI Data Release 1 Luminous Red Galaxy Sample} 
\author{J. Hou$^{1,2,3,4}$\,$^{\ast}$, R. N. Cahn$^{5}$, J.~Aguilar$^{5}$, S.~Ahlen$^{7}$, D.~Bianchi$^{8,9}$, D.~Brooks$^{10}$, T.~Claybaugh$^{5}$, P.~Doel$^{10}$, S.~Ferraro$^{5,12}$, J.~E.~Forero-Romero$^{13,14}$, E.~Gaztañaga$^{15,16,17}$, L.~Le~Guillou$^{29}$, G.~Gutierrez$^{18}$, K.~Honscheid$^{19,20,21}$, D.~Huterer$^{22,23}$, M.~Ishak$^{24}$, R.~Joyce$^{25}$, S.~Juneau$^{25}$, R.~Kehoe$^{26}$, D.~Kirkby$^{27}$, T.~Kisner$^{5}$, A.~Kremin$^{5}$, C.~Lamman$^{28}$, M.~Landriau$^{5}$, A.~de la Macorra$^{11}$, M.~Manera$^{30,31}$,  A.~de Mattia$^{6}$, R.~Miquel$^{31,32}$, E.~Mueller$^{33}$, S.~Nadathur$^{16}$, G.~Niz$^{34,35}$, W.~J.~Percival$^{36,37,38}$, F.~Prada$^{39}$, I.~P\'erez-R\`afols$^{40}$, A.~J.~Ross$^{19,21,41}$, G.~Rossi$^{42}$, E.~Sanchez$^{43}$, D.~Schlegel$^{5}$, M.~Schubnell$^{22,23}$, H.~Seo$^{44}$, J.~Silber$^{5}$, Z.~Slepian$^{4,5}$, D.~Sprayberry$^{25}$, G.~Tarl\'{e}$^{23}$, B.~A.~Weaver$^{25}$, H.~Zou$^{45}$}

\affiliation{$^1$Max-Planck-Institut f{\"u}r Extraterrestrische Physik, Postfach 1312, Giessenbachstrasse, 85748 Garching, Germany}
\affiliation{$^2$Institute of Astronomy, University of Cambridge, Madingley Rd, Cambridge CB3 0HA, UK}
\affiliation{$^3$Kavli Institute for Cosmology Cambridge, Madingley Road, Cambridge CB3 0HA, UK}
\affiliation{$^4$Department of Astronomy, University of Florida, Gainesville, FL 32611, USA}
\affiliation{$^5$Lawrence Berkeley National Laboratory, 1 Cyclotron Road, Berkeley, CA 94720, USA}
\affiliation{$^{6}$IRFU, CEA, Universit\`e Paris-Saclay, F-91191 Gif-sur-Yvette, France}
\affiliation{$^{7}$Department of Physics, Boston University, 590 Commonwealth Avenue, Boston, MA 02215 USA}
\affiliation{$^{8}$Dipartimento di Fisica ``Aldo Pontremoli'', Universit\`a degli Studi di Milano, Via Celoria 16, I-20133 Milano, Italy}
\affiliation{$^{9}$INAF-Osservatorio Astronomico di Brera, Via Brera 28, 20122 Milano, Italy}
\affiliation{$^{10}$Department of Physics \& Astronomy, University College London, Gower Street, London, WC1E 6BT, UK}
\affiliation{$^{11}$Instituto de F\'{\i}sica, Universidad Nacional Aut\'{o}noma de M\'{e}xico,  Circuito de la Investigaci\'{o}n Cient\'{\i}fica, Ciudad Universitaria, Cd. de M\'{e}xico  C.~P.~04510,  M\'{e}xico}
\affiliation{$^{12}$University of California, Berkeley, 110 Sproul Hall \#5800 Berkeley, CA 94720, USA}
\affiliation{$^{13}$Departamento de F\'isica, Universidad de los Andes, Cra. 1 No. 18A-10, Edificio Ip, CP 111711, Bogot\'a, Colombia}
\affiliation{$^{14}$Observatorio Astron\'omico, Universidad de los Andes, Cra. 1 No. 18A-10, Edificio H, CP 111711 Bogot\'a, Colombia}
\affiliation{$^{15}$Institut d'Estudis Espacials de Catalunya (IEEC), c/ Esteve Terradas 1, Edifici RDIT, Campus PMT-UPC, 08860 Castelldefels, Spain}
\affiliation{$^{16}$Institute of Cosmology and Gravitation, University of Portsmouth, Dennis Sciama Building, Portsmouth, PO1 3FX, UK}
\affiliation{$^{17}$Institute of Space Sciences, ICE-CSIC, Campus UAB, Carrer de Can Magrans s/n, 08913 Bellaterra, Barcelona, Spain}
\affiliation{$^{18}$Fermi National Accelerator Laboratory, PO Box 500, Batavia, IL 60510, USA}
\affiliation{$^{19}$Center for Cosmology and AstroParticle Physics, The Ohio State University, 191 West Woodruff Avenue, Columbus, OH 43210, USA}
\affiliation{$^{20}$Department of Physics, The Ohio State University, 191 West Woodruff Avenue, Columbus, OH 43210, USA}
\affiliation{$^{21}$The Ohio State University, Columbus, 43210 OH, USA}
\affiliation{$^{22}$Department of Physics, University of Michigan, 450 Church Street, Ann Arbor, MI 48109, USA}
\affiliation{$^{23}$University of Michigan, 500 S. State Street, Ann Arbor, MI 48109, USA}
\affiliation{$^{24}$Department of Physics, The University of Texas at Dallas, 800 W. Campbell Rd., Richardson, TX 75080, USA}
\affiliation{$^{25}$NSF NOIRLab, 950 N. Cherry Ave., Tucson, AZ 85719, USA}
\affiliation{$^{26}$Department of Physics, Southern Methodist University, 3215 Daniel Avenue, Dallas, TX 75275, USA}
\affiliation{$^{27}$Department of Physics and Astronomy, University of California, Irvine, 92697, USA}
\affiliation{$^{28}$Center for Astrophysics $|$ Harvard \& Smithsonian, 60 Garden Street, Cambridge, MA 02138, USA}
\affiliation{$^{29}$Sorbonne Universit\'{e}, CNRS/IN2P3, Laboratoire de Physique Nucl\'{e}aire et de Hautes Energies (LPNHE), FR-75005 Paris, France}
\affiliation{$^{30}$Departament de F\'{i}sica, Serra H\'{u}nter, Universitat Aut\`{o}noma de Barcelona, 08193 Bellaterra (Barcelona), Spain}
\affiliation{$^{31}$Institut de F\'{i}sica d’Altes Energies (IFAE), The Barcelona Institute of Science and Technology, Edifici Cn, Campus UAB, 08193, Bellaterra (Barcelona), Spain}
\affiliation{$^{32}$Instituci\'{o} Catalana de Recerca i Estudis Avan\c{c}ats, Passeig de Llu\'{\i}s Companys, 23, 08010 Barcelona, Spain}
\affiliation{$^{33}$Department of Physics and Astronomy, University of Sussex, Brighton BN1 9QH, U.K}
\affiliation{$^{34}$Departamento de F\'{\i}sica, DCI-Campus Le\'{o}n, Universidad de Guanajuato, Loma del Bosque 103, Le\'{o}n, Guanajuato C.~P.~37150, M\'{e}xico}
\affiliation{$^{35}$Instituto Avanzado de Cosmolog\'{\i}a A.~C., San Marcos 11 - Atenas 202. Magdalena Contreras. Ciudad de M\'{e}xico C.~P.~10720, M\'{e}xico}
\affiliation{$^{36}$Department of Physics and Astronomy, University of Waterloo, 200 University Ave W, Waterloo, ON N2L 3G1, Canada}
\affiliation{$^{37}$Perimeter Institute for Theoretical Physics, 31 Caroline St. North, Waterloo, ON N2L 2Y5, Canada}
\affiliation{$^{38}$Waterloo Centre for Astrophysics, University of Waterloo, 200 University Ave W, Waterloo, ON N2L 3G1, Canada}
\affiliation{$^{39}$Instituto de Astrof\'{i}sica de Andaluc\'{i}a (CSIC), Glorieta de la Astronom\'{i}a, s/n, E-18008 Granada, Spain}
\affiliation{$^{40}$Departament de F\'isica, EEBE, Universitat Polit\`ecnica de Catalunya, c/Eduard Maristany 10, 08930 Barcelona, Spain}
\affiliation{$^{41}$Department of Astronomy, The Ohio State University, 4055 McPherson Laboratory, 140 W 18th Avenue, Columbus, OH 43210, USA}
\affiliation{$^{42}$Department of Physics and Astronomy, Sejong University, 209 Neungdong-ro, Gwangjin-gu, Seoul 05006, Republic of Korea}
\affiliation{$^{43}$CIEMAT, Avenida Complutense 40, E-28040 Madrid, Spain}
\affiliation{$^{44}$Department of Physics \& Astronomy, Ohio University, 139 University Terrace, Athens, OH 45701, USA}
\affiliation{$^{45}$National Astronomical Observatories, Chinese Academy of Sciences, A20 Datun Road, Chaoyang District, Beijing, 100101, P.~R.~China}

\thanks{Email: \texttt{jiamin.hou@ast.cam.ac.uk}}

\begin{abstract}
We present a measurement of the non-Gaussian four-point correlation function (4PCF) from the DESI DR1 Luminous Red Galaxy (LRG) sample. For the gravitationally induced parity-even 4PCF, we detect a signal with a significance of 14.7$\sigma$ using our fiducial setup. We assess the robustness of this detection through a series of validation tests, including auto- and cross-correlation analyses, sky partitioning across multiple patch combinations, and variations in radial scale cuts. Due to the low completeness of the sample, we find that differences in fiber assignment implementation schemes can significantly impact estimation of the covariance and introduce biases in the data vector. After correcting for these effects, all tests yield consistent results. This is one of the first measurements of the connected 4PCF on the DESI LRG sample: the good agreement between the simulation and the data implies that
the amplitude of the density fluctuation inferred from the connected 4PCF is consistent with the Planck $\Lambda$CDM cosmology. The methodology and diagnostic framework established in this work provide a foundation for interpreting parity-odd 4PCF.
\end{abstract}


\maketitle

\section{Introduction}
\label{sec:intro}
Non-Gaussianity on cosmological scales can arise from late-time nonlinear gravitational evolution or from deviations in the initial conditions from the standard single-field, slow-roll inflation paradigm~\cite{Maldacena2003:PNG,Bartolo2004:PNG,Chen2010:PNG}. While two-point statistics can fully capture all information in a Gaussian random field, higher-order statistics offer an efficient framework for capturing this non-Gaussian information.

Beyond the standard two-point statistics, a variety of tools have been developed, such as $N$-point correlation functions in position space, and their Fourier-space counterparts, the polyspectra. The lowest-order extension beyond two-point statistics is the three-point statistics. In recent years, significant developments have been made for both the three-point correlation function (3PCF) and the bispectrum.~\cite{Szapudi2004:3PCF,Slepian201506,Scoccimarro2015:bkRSD,Sugiyama2018,DAmico2022:BOSSbispec,Philcox:2021kcw,Eggemeier2025:VDG}.
In addition to $N$-point functions, alternative statistics have been explored to extract non-Gaussian features, such as galaxy skew spectra~\cite{Schmittfull202103,Dizgah202004,Hou:2022rcd,Hou2024:SimbigSkewSpec}, density split statistics~\cite{Paillas2021:DensitySplit}, the wavelet scattering transform~\cite{Eickenberg:2022qvy,Blancard2023:simbigWST,Valogiannis2022:WST}, and marked two-point statistics~\cite{White2016:mark,Sheth2005:mark}.

While a range of statistics beyond the standard two-point function has been developed, most either focus on three-point statistics or extract additional non-Gaussian information \jhc{through nonlinear transformations}. As a result, they do not explicitly encode correlations among more than three fields. A key advantage of galaxy spectroscopic surveys is their sensitivity to the full three-dimensional galaxy spatial positions, making them naturally suited to probe symmetries such as parity, defined as spatial inversion. The sensitivity to the parity transformation can be explored using galaxy four-point functions~\cite{Cahn2021:parity,hou2022:parity,philcox_parity} or various compressed forms of four-point statistics~\cite{Jamieson2024:POP,Jeong2012:fossil}.

The galaxy four-point function can be decomposed into parity-even and parity-odd components. The even component itself contains both a disconnected part, composed of products of two-point functions; and a connected part arising from nonlinear gravitational evolution. The latter also encodes rich cosmological information in tightening cosmological parameters and understand primordial non-Gaussianity~\cite{Gualdi202104}. 

Previous work~\cite{Philcox2021boss4pcf} quantified the detection significance for the gravitationally induced connected even 4PCF using the galaxy sample from Baryon Oscillation Spectroscopic Survey (BOSS~\cite{Dawson2013}). In this paper, we study the galaxy sample from the Dark Energy Spectroscopic Instrument (DESI)~\cite{DESI2013:Snowmass,DESI2016:Science,DESI2016:InstrumentDesign,DESI2022:InstrumentOverview,DESI_DR1_2025,DESI_DR1_cosmology_2024,DESI_DR2_cosmology_2025,DESI_Survey_Operation_2023}, which currently provides the largest spectroscopic galaxy sample to date. In particular, we focus on the Luminous Red Galaxy (LRG) sample, enabling a direct comparison of information gain relative to BOSS. Moreover, this work is crucial for understanding systematics in the 4PCF and for interpreting the significance of parity-odd measurements.

The paper is structured as follows: in \S\ref{sec:desi_dr1} we provide an overview of the DESI DR1, the LRG sample, the simulations, and the evaluation of systematics, particularly those arising from sample incompleteness in the first-year data of the five-year survey program. 
In \S\ref{sec:4PCF_estimator}, we provide a review of the 4PCF estimator and the measurement on the real data. In \S\ref{sec:methodology} we discuss the methodology used in this paper in quantifying the detection significance. In \S\ref{sec:Result} we discuss the detection significance for the DR1 result.
In \S\ref{sec:discussion} we discuss various points regarding the scale dependence, usage of different covariance, impact of incompleteness. Finally, we conclude in \S\ref{sec:summary}.

Our Fourier transform  convention is $\tilde f(\bfk)=\int d^3x\, e^{-i\bfk\cdot \bfx} f(\bfx)$ and $f(\bfx) =\int_{\bfk} e^{i\bfk\cdot \bfx} \tilde{f}(\bfk)$, where $\int_{\bfk}\equiv \int d^3\bfk/(2\pi)^3$.

\section{An Overview of DESI DR1}
\label{sec:desi_dr1}
DESI is a ground-based, fiber-fed spectroscopic survey located at Kitt Peak National Observatory in Arizona~\cite{DESI_Fiber_System_2024,DESI_spectroscopic_pipeline_2023}. Over its five-year duration, DESI will map 14,200 square degrees of the sky, aiming to collect spectra for approximately 40 million galaxies and quasars across the redshift range $0.1 < z < 4.2$~\cite{Myers2023:DESItargetSelection}. This includes the Bright Galaxy Survey (BGS) targeting galaxies at $0.1 < z < 0.4$~\cite{Hahn2023:DESIBGSselection}. During dark time, DESI observes luminous red galaxies (LRGs) in the range $0.4 < z < 1.1$~\cite{Zhou2023:DESILRGselection}, emission-line galaxies (ELGs) at $0.8 < z < 1.6$~\cite{Raichoor2023:DESIELGselection}, and quasars (QSOs) spanning $0.8 < z < 2.1$~\cite{Chaussidon2023:DESIQSOselection}. DESI uses robotic positioners to place optical fibers within the 7-square-degree field of view of its focal plane~\cite{Miller2024:DESIOpticalCorrector}. Each set of targets is assigned to a set of 5,000 fibers. The year-one (Y1) data were obtained during the first
13 months of main survey operation. Over 4.7 million objects covering a sky area of 7,500 square degrees are included in the clustering analyses~\cite{DESICollaboration2024:FullShape}. For this work, we focus on a sample of 2,138,627 LRGs, which we briefly describe in the following section.

\subsection{DESI DR1 LRG Sample}

All four classes of DESI targets, including the LRGs, were selected based on photometry from Data Release 9 (DR9) of the Legacy Survey (LS) imaging~\cite{Dey2019:DESILS}. 
The Legacy Survey combines photometric data from the Beijing-Arizona Sky Survey (BASS~\cite{Zou2017:BASS}) and the Mayall z-band Legacy Survey (MzLS) for target selection in the Northern Galactic Cap (NGC). For the Southern Galactic Cap (SGC), target selection is based on imaging from the Dark Energy Camera (DECam~\cite{Flaugher2015:DECam}) and the Dark Energy Survey (DES~\cite{DES2005}). Infrared photometry in the W1 and W2 bands from the WISE satellite~\cite{Wright2010:WISE,Lang2014:unWISE} is used across the entire sky. 

The DESI LRG sample is selected using g,r,z, and  W1 flux measurements~\cite{Zhou2023:DESILRGselection}. A series of photometric cuts is applied to the sample to reduce stellar contamination, select the most luminous galaxies across redshifts, maintain an approximately constant comoving number density in the range $0.4 < z < 0.8$, and ensure a high spectroscopic-redshift success rate.
The selection criteria are independently optimized in the BASS/MzLS and DECam regions to yield a sample of passively evolving galaxies with an approximately constant number density of $5 \times 10^{-4} \,[\hmpc]^{3}$ in the redshift range $0.4 < z < 0.8$. Beyond this range, the number density drops below $1 \times 10^{-4} \,[\hmpc]^{3}$ by $z = 1.1$, primarily due to a threshold on the $z$-band fiber magnitude~\cite{Zhou2023:DESILRGselection}. 

To obtain galaxy spectra, each target is assigned a fiber that directs its emitted light into a DESI spectrograph. A \textsc{fiberassign} code~\cite{Lasker2025:DESIFiberAssign} is used to allocate fibers to designated targets based on the DESI targeting algorithm. This algorithm assigns fibers according to the merged-target-ledger (MTL) file, which records sample information such as sky coordinates, target IDs, and priorities. Among these, the priority label is particularly relevant: in this process, the LRG sample has intermediate priority, lower than that of the QSO sample but higher than that of the ELG sample. 

Fiber assignment in DESI has two major impacts. First, each fiber can access only a limited patrol area; although adjacent patrol regions overlap and about 15\% of the focal plane is reachable by two fibers, in dense regions the number of targets often exceeds the number of available fibers. Second, due to the physical size of the positioners, fibers cannot be placed too close together, a constraint commonly referred to as ``fiber collisions'', which limits the ability to observe all targets, especially for closely spaced galaxy pairs in a single pass. Each pointing of the telescope brings approximately 25,000 targets into the focal plane, of which only 5,000 can be accommodated.  As the survey progresses, each domain is visited multiple times.  Early in the survey there is significant incompleteness. In the full five-year survey, these limitations will be largely mitigated by repeated observations.

In addition to fiber assignment incompleteness, other systematics such as density variation due to imaging systematics, redshift failure also need to taken into account. These effects are currently accounted for by applying weights to either data or randoms at catalog level.

\begin{figure*}
    \centering
    \includegraphics[width=0.9\linewidth]{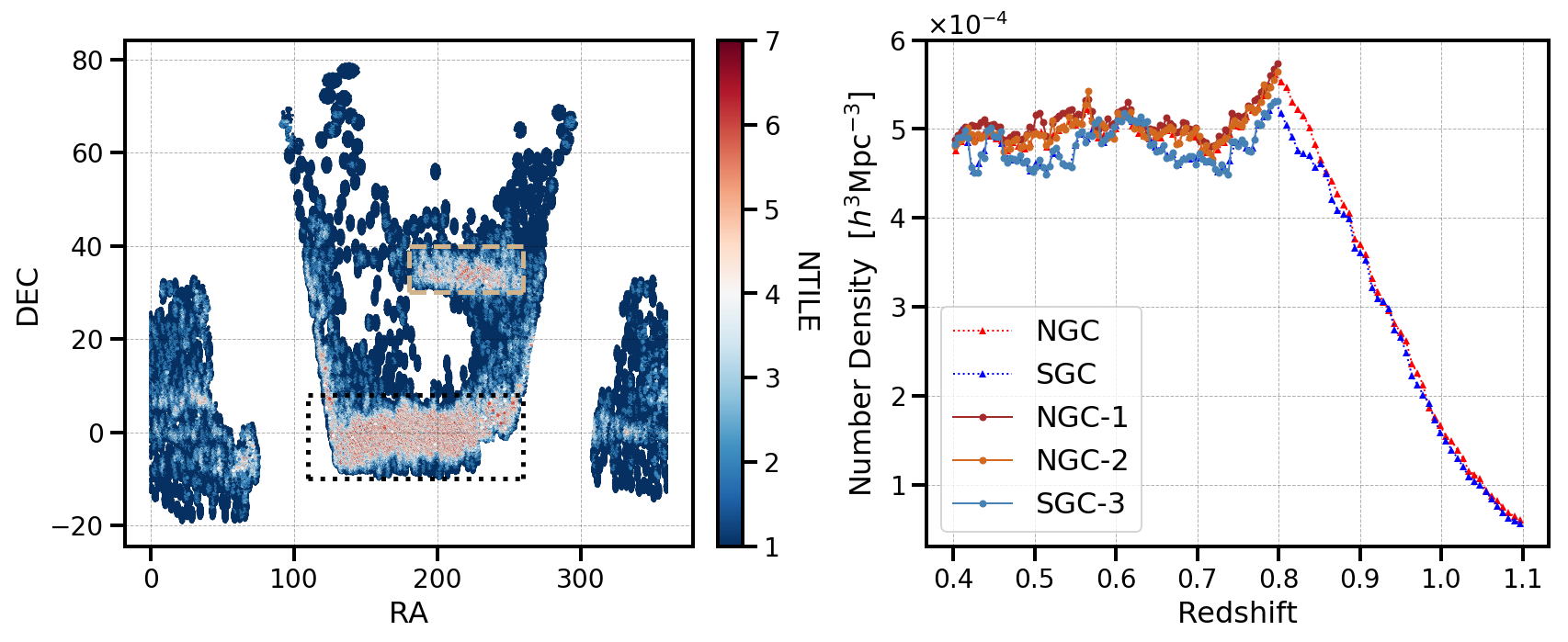}
    \caption{{\it Left}: footprint of DESI-DR1 LRG. The color map shows the number of targets sharing the same unique identifier, which combines the tile ID and fiber location. This quantity is closely correlated with the sample completeness: higher values of \textsc{NTILE} indicate higher completeness. For this paper, we specifically studied two regions, the ${\rm NGC}_1$ (dotted black box) and ${\rm NGC}_2$ (dashed gray box). {\it Right}: number density as a function of redshift $n(z)$ for the NGC (red) and SGC (blue), including different patches with various cuts in both angular and radial directions. After applying completeness weights, the number densities as a function of redshift are consistent across various cuts.  This demonstrates the effectiveness of the completeness weights at the density level in the radial direction.}
    \label{fig:footprint_n_of_z_NGCxSGC}
\end{figure*}

\subsection{Choice of the LRG Sample}
\label{sec:sample_selection}
A key consideration for the DR1 LRG sample is its completeness, defined as the ratio of observed galaxies to the targeted galaxies for a given region. 
Fig.~\ref{fig:footprint_n_of_z_NGCxSGC} shows the DESI DR1 LRG footprint, where the colormap indicates the number of targets sharing the same identifier (\textsc{TILELOCID}, which combines the tile ID and fiber location). The quantity \textsc{NTILE} denotes the number of tiles on which a target was observed, serving as an indicator of sample completeness: higher values correspond to more complete regions~\cite{DESI2024:SampleDefinition}.

We use the default clustering catalog with a redshift range $0.4 < z < 1.1$. Additionally, given the dependence on completeness and redshift evolution, we further split the LRG sample into three subregions. For the NGC, we select areas with higher completeness, defining two subregions: NGC-1 with angular cuts in both right ascension (RA) and declination (DEC): $110<\text{RA}<260$, $-10<\text{DEC}<8$, and a radial cut $0.4<z<0.8$; and NGC-2 with $180<\text{RA}<260$, $30<\text{DEC}<40$, and the same redshift range. For the SGC, we define a single subregion, SGC-3, using only the radial cut $0.4<z<0.8$, since there is no large variation in completeness (unlike NGC) and also to retain statistical constraining power. The geometric selection, the number of galaxies, and the corresponding volumes are listed in Table~\ref{tab:statistic}.

\begin{table*}[t]
\centering
\caption{Statistics for the DR1 LRG sample, including the number of galaxies $N_{\rm gal}$; both the number density $\bar{n}_{\rm g}$ and fiducial volume $V_{\rm fid}$ are obtained from fitting the analytic covariance matrix. The effective volume $V_{\rm eff}$ is computed from the trace of the product of the inverse of the full analytic covariance and the covariance for the $\mu$-th patch ({\it c.f.}~\S\ref{sec:cross-corr}). The difference between $V_{\rm fid}$ and $V_{\rm eff}$ arises from variations in number density.}
\label{tab:statistic}
\begin{tabular}{@{}p{3cm}p{3cm}p{3cm}p{3cm}p{3cm}p{3cm}@{}}
\toprule
Region & Selection & $N_{\rm gal}$ & $\bar{n}_{\rm g}~[\mpch]^{-3}$ & $V_{\rm fid}~[\gpch]^3$ & $V_{\rm eff}~[\gpch]^3$ \\ \midrule
Full-NGC & -- & 1,476,135 & $2.3\times 10^{-4}$ & 2.54 & 2.54 \\ [3ex]
$\quad$ NGC-1 & \begin{tabular}[c]{@{}l@{}}$110 <\text{RA} < 260$, \\ $-10 <\text{DEC} < 8$, \\ $0.4<z<0.8$ \end{tabular} & 434,054 & $3.5\times 10^{-4}$ & 0.54 & 0.87 \\[5ex]
$\quad$ NGC-2 & \begin{tabular}[c]{@{}l@{}}$180 <\text{RA} < 260$, \\ $30 <\text{DEC} < 40$, \\ $0.4<z<0.8$\end{tabular} & 119,034 & $2.9\times 10^{-4}$ & 0.13 & 0.17 \\ \midrule
Full-SGC & -- & 662,492 & $1.5\times 10^{-4}$ & 1.43 & 0.79 \\ [3ex]
$\quad$ SGC-3 & $0.4<z<0.8$ & 398,089 & $1.7\times 10^{-4}$ & 0.72 & 0.48 \\ \bottomrule
\end{tabular}
\end{table*}

\section{Simulations}
\label{sec:simulations}
Two types of simulations are used for DESI DR1 for estimating the covariance matrix and studying systematics. 

To estimate the covariance matrix, we use a suite of fast, approximate, but extensive  \ezmock~\cite{Chuang2015:EZmock,Zhao2021:EZmock} provided by the DESI collaboration. For the \ezmock, the dark matter density field is based on the Zel’dovich
approximation. The resulting density field is populated with galaxies using an
effective-bias model to account for non-linear evolution and galaxy bias. There are 1000 realizations for the box of size $L_{\rm box}=6\,\gpch$, which are post-processed to match the DESI DR1 geometry.

In addition, DESI also has a suite of high-resolution $N$-body simulations: \textsc{AbacusSummit}~\cite{Garrison2019:Abacus,Maksimova2021:Abacus}.  These  have $L_{\rm box}=2\,\gpch$ box and use a fiducial cosmology $\Omega_{\rm c} h^2=0.1200$,  $\Omega_{\rm b} h^2=0.02237$, $\sigma_8=0.811355$, $n_s=0.9649$, $h=0.6736$, $w_0=-1$, $w_a=0$. They use a  halo occupation distribution model for the galaxy-halo connection, calibrated on the DESI Early Data Release (EDR~\cite{Adame2024:DESIEDR}) described in~\cite{Yuan2024:DESIAbacusHODLRG} for the LRG sample. Since the mock box size is insufficient to encompass the full DESI DR1 LRG volume, the mocks are tiled to fill the survey footprint. As a result, they overestimate  the cosmic variance and are therefore unsuitable for covariance estimation. Nevertheless, they remain valuable for studying systematics.

Both \ezmock\ and \abacus\ simulations are post-processed with three variations for the fiber assignment implementation. The ``complete'' mocks  represent surveys capturing all galaxies that could have been targeted~\footnote{For the \abacus-Complete mocks, we additionally downsample it to match the number density of the altMTL mocks, so that their statistical fluctuations are comparable as those in the real data}. The ``alternative MTL'' (altMTL) mocks represent the most realistic fiber assignment: they were generated using the DESI \textsc{fiberassign} code~\cite{Lasker2025:DESIFiberAssign} applied to tiles in the same ordering and cadence, with a feedback loop to the target list, mirroring the procedure used for the observed data. The ``fast fiber assignment'' (FFA) mocks emulate the fiber assignment process by repeatedly sampling from the average targeting probability of galaxies. This probability is learned from the data as a function of the number of overlapping tiles and the local angular clustering.

We will discuss in more detail the fiber assignment implementations and their impact on statistical fluctuations of the 4PCF (see Sec.~\ref{sec:incomplete_stat_fluctuations}), which is crucial for interpreting the detection significance. 

\section{Four-point Correlation Function and its Estimator}
\label{sec:4PCF_estimator}
Assuming homogeneity, the four-point correlation can be expressed as $\zeta(\bfr_1,\bfr_2, \bfr_3)$ by taking the origin at one galaxy. It can be decomposed into a parity-even part $\zeta_+$ and a parity-odd part $\zeta_-$
\begin{eqnarray}\label{eqn:4PCF_even_odd}
\zeta(\bfr_1,\bfr_2, \bfr_3) = \zeta_{+}(\bfr_1,\bfr_2, \bfr_3) + i\zeta_{-}(\bfr_1,\bfr_2, \bfr_3),
\end{eqnarray}
where $\zeta_+$ and $\zeta_-$ are real.
The parity-odd part in general vanishes for the standard model of cosmology (further discussion see~\cite{Cahn2021:parity,hou2022:parity,philcox_parity}). The parity-even part includes both a disconnected contribution, given by products of two-point functions $\xi$ and a connected contribution $\zeta^{(c)}_{+}$.  If initially the distribution is given by a Gaussian random field, there is only a disconnected piece.  Non-linear gravitation evolution generates the connected piece:
\begin{eqnarray}\label{eqn:connected_4pcf}
&&\zeta_{+}(\bfr_1,\bfr_2, \bfr_3)\nonumber\\
&&\qquad = \zeta^{(c)}_{+}(\bfr_1,\bfr_2, \bfr_3) +\left[ \xi(\bfr_1)\xi(\bfr_2 - \bfr_3) + \text{2 cyc.}\right],
\end{eqnarray}
where we cyclically permute the contribution from the 2PCF.

\subsection{Isotropic Basis Functions}
A brute force measurement of the 4PCF scales as $\calO(N_g^4)$, with $N_g$ being the number of galaxies and is thus computationally challenging. To accelerate the process, we expand the 4PCF in a basis of isotropic functions $\mathcal{P}_{\ell_1 \ell_2 \ell_3}$ \cite{Cahn202010}, which capture its angular behavior about one galaxy (the ``primary''), multiplied by functions that capture its dependence on tetrahedron side lengths $r_1, r_2, r_3$ from that primary. 
The radial coefficient function are projections of the full distribution on the isotropic functions:
\begin{align}
    &\zeta_{\ell_1\ell_2\ell_3}(r_1,r_2,r_3) = \\
    &\int d\hr_1\,d\hr_2\,d\hr_3\, {\zeta}(\bfr_1,\bfr_2,\bfr_3) \,\mathcal{P}^*_{\ell_1 \ell_2 \ell_3}(\hat{r}_1, \hat{r}_2, \hat{r}_3),\nonumber
\end{align}
where $\ell_i$, for $i=1,2 ,3$ are the ``angular momenta'' associated with the three direction vectors $\bfr_i$, and star denotes a complex conjugate. We define the angular differential $d\hr$ by $\int d\hr = 4\pi$.

To avoid an over-complete basis,we expand in $\,\mathcal{P}^*_{\ell_1 \ell_2 \ell_3}(\hat{r}_1, \hat{r}_2, \hat{r}_3)$ where  $r_1 < r_2  < r_3$. The isotropic functions $\mathcal{P}_{\ell_1 \ell_2 \ell_3}(\hat{r}_1, \hat{r}_2, \hat{r}_3)$ of three arguments is given by
\begin{align}
\label{eqn:Plll}
\mathcal{P}_{\ell_1 \ell_2 \ell_3}(\hat{r}_1, \hat{r}_2, \hat{r}_3) &= \sum_{m_1 m_2 m_3} 
(-1)^{\ell_1+\ell_2+\ell_3}\six{\ell_1}{\ell_2}{\ell_3}{m_1}{m_2}{m_3}\nonumber\\
&\quad \times Y_{\ell_1 m_1}(\hat{r}_1)
Y_{\ell_2 m_2}(\hat{r}_2)
Y_{\ell_3 m_3}(\hat{r}_3),
\end{align}
where the $Y_{\ell_i m_i}(\hr_i)$ are spherical harmonics and the $m_i$ are the $z$-components of their angular momenta $\ell_i$. The product of the three spherical harmonics is weighted by the Wigner 3-$j$ symbol. The isotropic basis is fully separable in the $\hr_i$, thus reducing the formal scaling of the 4PCF estimator to pair-wise operations~\footnote{In the large-$N_{\rm g}$ limit, with typical number of galaxies of modern galaxy surveys reaches $\calO(10^6)$, the algorithm scales linearly with the number of objects. This is because the computational cost is dominated by assembling all galaxy quartets for each primary galaxy, whereas the computation of the harmonic expansion coefficients is sub-dominant (see also~\cite{Philcox:encore}).}. 

Applying a parity transformation $\mathbb{P}\colon (x,y,z) \rightarrow (-x,-y,-z)$ to a spherical harmonic yields $\mathbb{P}[Y_{\ell m}(\hat{r})] = (-1)^{\ell} Y_{\ell m}(\hat{r})$. Extending this relation to the isotropic basis function in Eq.~\eqref{eqn:Plll}, we find:
\begin{eqnarray}
    &&\mathbb{P}\left[\mathcal{P}_{\ell_1 \ell_2 \ell_3}(\hat{r}_1, \hat{r}_2, \hat{r}_3)\right]=(-1)^{\ell_1+\ell_2+\ell_3}\mathcal{P}_{\ell_1 \ell_2 \ell_3}(\hat{r}_1, \hat{r}_2, \hat{r}_3).
\end{eqnarray}
With this basis, the parity behavior described in Eq.~\eqref{eqn:4PCF_even_odd} can be isolated by examining the sum $\sum_i\ell_i$: the basis function is parity-even if the sum is even, and parity-odd if the sum is odd.

In practice, the connected 4PCF multipole estimator $\hat\zeta^{\rm (c)}_{\ell_1\ell_2\ell_3}$ is obtained by subtracting from the full 4PCF multipole estimator $\hat\zeta_{\ell_1\ell_2\ell_3}$  the  disconnected 4PCF estimator $\hat\zeta^{\rm (dc)}_{\ell_1\ell_2\ell_3}$ as follows:
\begin{eqnarray}
\hat\zeta^{\rm (c)}_{\ell_1\ell_2\ell_3}(r_1,r_2,r_3) = \hat\zeta_{\ell_1\ell_2\ell_3}(r_1,r_2,r_3) - \hat\zeta^{\rm (dc)}_{\ell_1\ell_2\ell_3}(r_1,r_2,r_3),
\end{eqnarray}
where the disconnected multipoles arise from the product of two two-point correlation functions (2PCFs, {\it c.f}, Eq.~\ref{eqn:connected_4pcf}): one depending on a single argument, and the other on a composite argument. After locally projecting both 2PCFs onto the spherical harmonic basis, the resulting estimators are summed, yielding global 2PCF multipoles with single angular momentum indices $\xi_{\ell m}$, and double indices $\xi_{\ell m, \ell' m'}$. The details of measuring the connected 4PCF and definition of the 2PCF multipoles $\xi_{\ell m}$ and $\xi_{\ell m, \ell' m'}$ are given in~\cite{Philcox2021boss4pcf}. 

Hereafter, we will use $\zeta$ to denote the 
connected parity-even 4PCF multipoles and drop all the superscripts and subscripts for brevity. 

\subsection{Survey Geometry Correction}
\label{sec:survey_geometry}
The survey geometry can affect galaxy clustering measurements. To account for this, a random catalog with identical survey geometry is used as a correction. For a discrete galaxy sample, the galaxy number density can be expressed as
\begin{eqnarray}
    n_{\rm g, r}(\bfx)  = \sum_{i=1}^{N_{\rm g}} w^{\rm{g},\rm{r}}_i \delta_{\rm D}^{[3]}\left(\bfx-\bfx_i\right) , 
\end{eqnarray}
with $\{\rm{g},\rm{r}\}$ denoting the galaxy or random catalog, $w_i$ is the weight for each galaxy, $\delta_{\rm D}^{[3]}\left(\bfx-\bfx_i\right)$ is the 3-dimensional Dirac delta function. The number density can be converted to number counts $D(\bfx)=n_{\rm g}(\bfx)\Delta V$ and $R(\bfx)=n_{\rm r}(\bfx)\Delta V$, where $\Delta V$ is a volume element, which cancels out when taking the ratio. 
By forming the difference between the data and the random catalog, we can construct $N(\bfx) = D(\bfx) - R(\bfx)$. The extended Landy-Szalay estimator for the 4PCF thus reads~\cite{SE_3pt, Philcox:encore}. 

\begin{eqnarray}\label{eqn:extended_LS}
    \hat{\zeta}(\bfr_1,\bfr_2,\bfr_3) 
    &=& \frac{\int d^3\bfx\,N(\bfx)N(\bfx+\bfr_1)N(\bfx+\bfr_2)N(\bfx+\bfr_3)}{\int d^3\bfx\,R(\bfx)R(\bfx+\bfr_1)R(\bfx+\bfr_2) R(\bfx+\bfr_3)} \nonumber\\
    &=&\frac{\mathcal{N}(\bfr_1,\bfr_2,\bfr_3)}{\mathcal{R}(\bfr_1,\bfr_2,\bfr_3)}.
    \label{eqn:est}
\end{eqnarray}
Since we are interested in the coefficients of the 4PCF in the isotropic basis, we expand both sides of the equation in this basis. In particular, both the numerator and denominator on the right-hand side are also expanded in the isotropic basis. By doing so, Eq.~\eqref{eqn:extended_LS} becomes
\begin{eqnarray}
\hat{\zeta}_{\ell_1\ell_2\ell_3}(r_1,r_2,r_3) &=& \sum_{\ell'_1\ell'_2\ell'_3} \left[{\rm {\bf M}}^{-1}\right]_{\ell_1\ell_2\ell_3, \ell'_1\ell'_2\ell'_3}(r_1,r_2,r_3)\,\nonumber\\ && \, \times\,\frac{\calN_{\ell'_1\ell'_2\ell'_3}(r_1,r_2,r_3)}{\calR_{000}(r_1,r_2,r_3)}.
\label{eqn:4pcf_estimator_edge_corrected}
\end{eqnarray}
The mode decoupling matrix $\bf{M}$ reads~\cite{Philcox:encore}
\begin{eqnarray}
&&{\rm \bf{M}}_{\ell_1\ell_2\ell_3, \ell'_1\ell'_2\ell'_3}(r_1,r_2,r_3)\nonumber\\
&=& (4\pi)^{-3/2} (-1)^{\ell'_1+\ell'_2+\ell'_3} \sum_{L_1L_2L_3}\;\frac{\calR_{L_1L_2L_3}(r_1,r_2,r_3)}{\calR_{000}(r_1,r_2,r_3)} \nonumber\\
&&\,\times\,\prod_{i=1}^3 D_{\ell_iL_i\ell'_i}\; \calC^{\ell_i L_i \ell'_i }_{000}\nine{\ell_1}{L_1}{\ell'_1}{\ell_2}{L_2}{\ell'_2}{\ell_3}{L_3}{\ell'_3}.
\label{eqn:4pcf_coupling_matrix}
\end{eqnarray}
Here we use the Wigner 9-$j$ symbol and define the coefficients
\begin{eqnarray}\label{eqn:calD_l1l2l3}
\calD_{\ell_1\ell_2\ell_3} = \sqrt{(2\ell_1+1)(2\ell_2+1)(2\ell_3+1)}
\label{eqn:DP}
\end{eqnarray}
and
\begin{eqnarray}\label{eqn:calC_l1l2l3}
\calC^{\ell_1\ell_2\ell_3}_{000}\equiv \six{\ell_1}{\ell_2}{\ell_3}{0}{0}{0}.
\end{eqnarray}

\subsection{Measurement of the Connected 4-Point Correlation Functions}
We apply the 4PCF estimator described in this section to the DESI DR1 LRG sample~\footnote{We use the GPU version of the public code~\url{https://github.com/oliverphilcox/encore} with extensions including $\ell_{\rm max}$ for the 4PCF measurement}. For each galaxy, we apply weights following~\cite{DESI2024:SampleDefinition}:
\begin{eqnarray}\label{eqn:weights}
    w_{\rm tot} = w_{\rm comp} w_{\rm sys} w_{\rm zfail} / \av{w_{\rm comp}}_{N_{\rm tile}},
\end{eqnarray}
where $w_{\rm comp}$ is the completeness weight, defined as the number of targets that compete for a given fiber. $w_{\rm sys}$ accounts for target density fluctuations due to imaging conditions, including stellar density, HI column density, $z$-band galaxy depth, $r$-band PSF size, and $W1$ PSF depth, and is derived using a linear regression method~\cite{DESI2024:SampleDefinition}.  $w_{\rm zfail}$ accounts for the redshift success rate and reduces spurious fluctuations in tracer density that are correlated with observational conditions in DESI spectra. $\av{w_{\rm comp}}_{N_{\rm tile}}$ is the averaged completeness for a given number of overlapping tiles $N_{\rm tile}$.

Additionally, we apply FKP weights ~\cite{Feldman1994}, which balance the trade-off between cosmic variance and the Poissonian shot noise:
\begin{eqnarray}
    w_{\rm FKP} = \Big(1+n_0\av{C_{\rm assign}}_{\rm N_{\rm tile}}P_0\Big)^{-1},
\end{eqnarray}
where we take $n_0=4\times 10^{-4}\, (\hmpc)^3$ and $P_0=10,000\,(\mpch)^3$, and $\av{C_{\rm assign}}_{\rm N_{\rm tile}}$ is the expectation value of assignment completeness for a given number of overlapping tiles.

\begin{figure*}
    \centering
    \includegraphics[width=.96\linewidth]{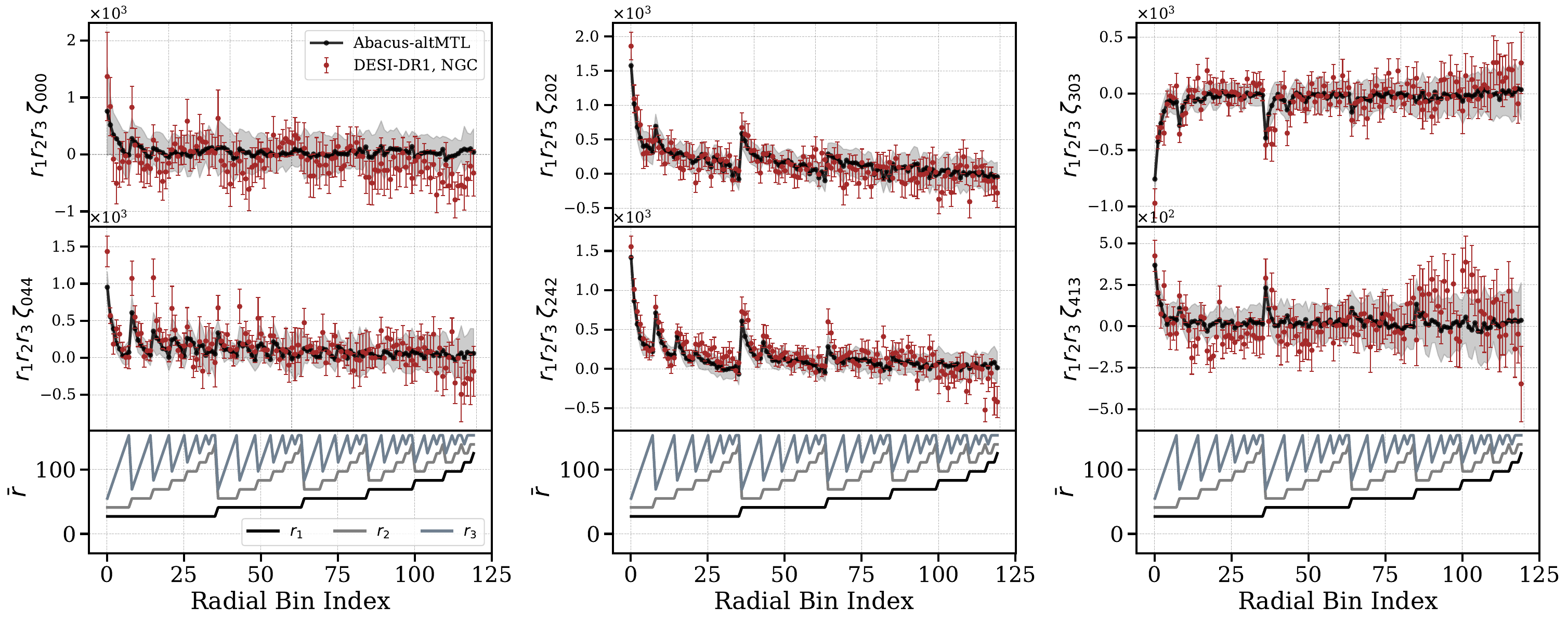}
    \caption{Measurement of the connected 4PCFs using the DESI-DR1 LRG sample in NGC. The top two panels show a subset of the angular channels ($\ell_1,\ell_2,\ell_3$) for the coefficients of the connected 4PCF, weighted by $r_1r_2r_3$. Points with error bars (red) are the measurement from the data; the black curves with shaded gray regions indicate the $1\sigma$ standard deviation from the Abacus simulations with altMTL fiber assignment scheme. The bottom panel shows the arrangement of the three radial bins.}
    \label{fig:4PCF_Iron_AbacusAltmtl_NGC_6ells}
\end{figure*}

\begin{figure*}
    \centering
    \includegraphics[width=.96\linewidth]{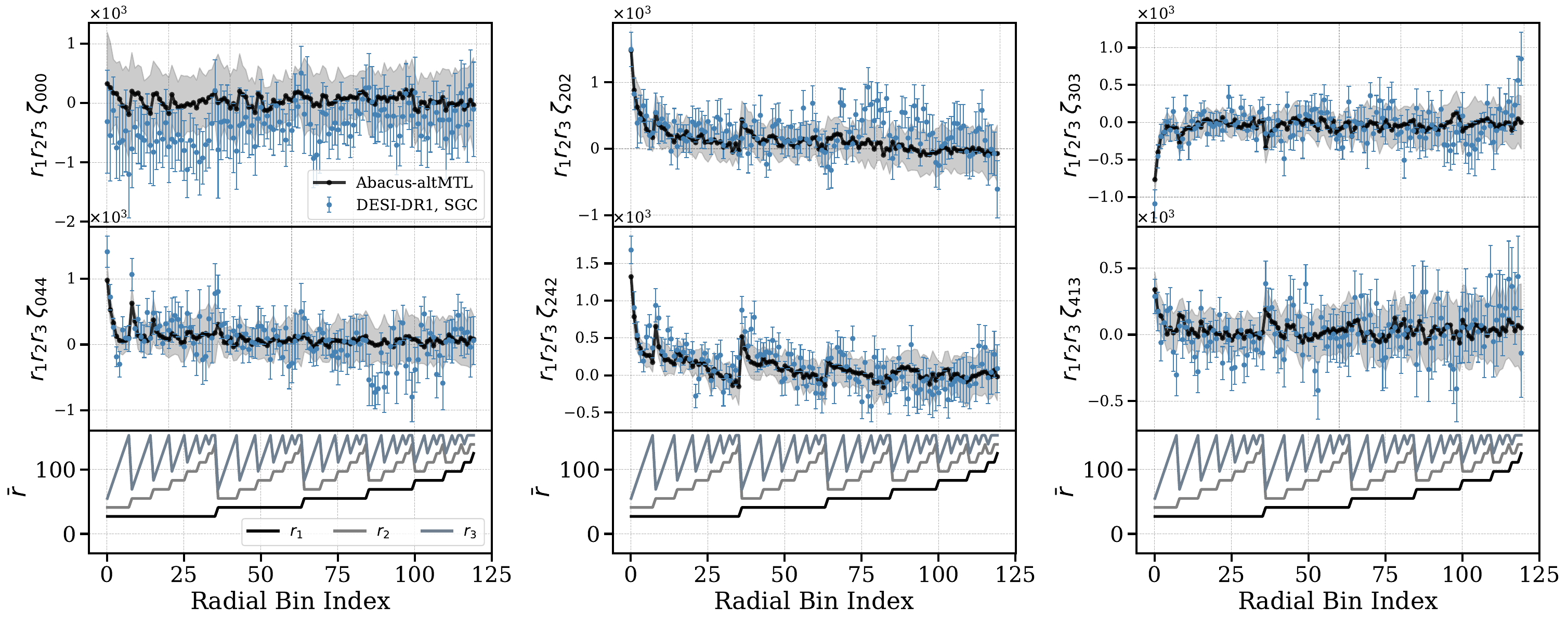}
    \caption{Simlar to Fig.~\ref{fig:4PCF_Iron_AbacusAltmtl_NGC_6ells}, this figure shows the measurement of the connected 4PCFs using the DESI-DR1 LRG sample in SGC. The top two panels show a subset of the angular channels ($\ell_1,\ell_2,\ell_3$) for the coefficients of the connected 4PCF, weighted by $r_1r_2r_3$. Points with error bars (blue) are the measurement from the data; the black curves with shaded gray regions indicate the $1\sigma$ standard deviation from the Abacus simulations with altMTL fiber assignment scheme. The bottom panel shows the arrangement of the three radial bins.}
    \label{fig:4PCF_Iron_AbacusAltmtl_SGC_6ells}
\end{figure*}

Fig.~\ref{fig:4PCF_Iron_AbacusAltmtl_NGC_6ells} and Fig.~\ref{fig:4PCF_Iron_AbacusAltmtl_SGC_6ells} show the measurement of the connected 4PCF using DESI-DR1 LRG sample in the NGC and SGC, respectively. We use the radial scales between $20\,\mpch < r < 160\,\mpch$, divided into ten bins. The top two panels show, for selected choices of $(\ell_1,\ell_2,\ell_3)$ the coefficients of the connected 4PCF, weighted by the product of $r_1r_2r_3$. Points with error bars are the measurement from the data; the black curves with shaded gray regions indicate the $1\sigma$ standard deviation from the Abacus simulations with altMTL fiber assignment scheme. The bottom panel shows the arrangement of the three radial bins. \jhc{The radial bin index can be expressed as $i = i_1(n-i_2+1) + (i_2-1)$, with $i_1$ and $i_2$ being the first and second bin index, and $n$ being the total number of radial bins.}

We select these angular channels because they exhibit prominent features, notably a characteristic ``sawtooth'' pattern. We note that these features are closely related to the specific ordering of the three sides $r_1$, $r_2$, $r_3$, where the signal-to-noise ratio is highest when the three bins are closest to each other and on small scales. 

\section{Analysis Methodology}
\label{sec:methodology}
To quantify the gravity-induced connected 4PCF,
we perform the analysis using the null hypothesis. With the assumption that the gravity-induced connected 4PCF is zero we assess the extent to which this hypothesis can be rejected. To do so, we compress the 4PCF data vector into a single scalar quantity, $\chi^2$, which involves the data vector, a model, and a covariance matrix.

The key challenge is accurate estimation of the covariance matrix. A common practice is to use simulations to estimate the covariance matrix.
The simulations used for DESI DR1 are, however, not tuned to match higher-order clustering, and the approximate method may underestimate nonlinear effects from gravitational dynamics, leading to a data-simulation mismatch. Additionally, the sample covariance requires the number of simulations  $N_{\rm sim}$  to be much larger than the length of the data vector. However, given that the length of the data vector is $\calO{(10^3)}$, alternative covariance estimation methods are necessary. 

\jhc{In \S\ref{sec:analyt-cov}, we present the formalism for computing the analytic covariance matrix following~\cite{Hou2022:AnalytCov}. This approach relies on several assumptions, including that the underlying field is a Gaussian random field and that survey geometry effects are neglected. To account for these limitations, we adopt a hybrid strategy: we use the analytic covariance to decompose the sample covariance estimated from simulations. This allows us to efficiently reduce the dimensionality of the data vector.}

\jhc{However, there remains a potential concern that the covariance estimated from simulations may not fully capture the true fluctuations present in the data.} To address the data-simulation mismatch \jhc{in the covariance}, we perform two tests: (1) auto-correlation  (\S\ref{sec:auto-corr}) and (2) cross-correlation (\S\ref{sec:cross-corr}). \jhc{The auto-correlation captures both the signal and statistical fluctuations, while the cross-correlation enables separation of these components. If the excess in $\chi^2$ from the auto-correlation arises from an underestimated covariance in the simulation, the cross-correlation will yield a reduced significance. Comparing the two provides a cross-check on the robustness of the detection significance.}

\subsection{Auto-correlation}
\label{sec:auto-corr}
We define $\chi^2$, which quantifies the \textit{overall} deviation from the null hypothesis for the parity-even signal and corresponds to the square of the detection significance (in units of $\sigma$)~\footnote{Here we apply the Einstein summation notation.}:\

\begin{eqnarray}
\label{eqn:chi2-auto}
    \chi^2 = \hat\zeta^{i} \mathbb{\hat C}^{-1}_{ij} \hat\zeta^{j}.
\end{eqnarray}
$\mathbb{\hat C}$ is an estimator of the covariance matrix, which can be derived analytically (see \S\ref{sec:analyt-cov}) or from simulations. $\hat\zeta$ is the measured 4PCF from the data, which is composed of a gravity-induced connected 4PCF signal $\zeta_{\rm g}$. The components of the data vector -- the values of $\zeta_{(\ell_1,\ell_2,\ell_3)}(r_1,r_2,r_3)$ -- are labeled $\zeta^i$. Following~\citep{Philcox2021boss4pcf,hou2022:parity}, we set $\zeta_{\rm g}$ to be zero for the null hypothesis test. 

We assume that the observed connected 4PCF is composed of the following terms
\begin{eqnarray}\label{eqn:zeta_3_term}
    \hat\zeta = \zeta_{\rm g} + \hat\zeta_{\rm s} + \hat\epsilon,
\end{eqnarray}
where $\zeta_{\rm g}$ represents the true gravitationally-induced signal, while $\hat\zeta_{\rm s}$ accounts for systematic effects, including survey systematics and inaccuracies in modeling nonlinear gravitational effects. The third term $\hat{\epsilon}$ corresponds to statistical fluctuations arising from cosmic variance. 

Broadly, systematic effects can impact the signal in two ways. On the one hand, they may introduce a non-vanishing bias, on the other hand, they can have non-vanishing correlations: $\hat\zeta_{\rm s}=\zeta_{\rm s}+\hat{\epsilon}_{\rm s}$, with $\zeta_{\rm s}$ being a systematic bias that survives in the ensemble average, and the stochastic $\hat\epsilon_{\rm s}$ term introduces a correction in addition to the cosmic variance:
\begin{eqnarray}
    \av{\hat\zeta_{\rm s}} = \zeta_{\rm s} \neq 0,\qquad \av{\hat\epsilon_{{\rm s}}^i \hat\epsilon_{{\rm s}}^j} \equiv \Delta\mathbb{C}_{\rm ss}^{ij}\neq 0
\end{eqnarray}
The statistical fluctuation becomes zero under the ensemble average, and has non-vanishing correlation:
\begin{eqnarray}\label{eqn:epsilon_cs}
    \av{\hat\epsilon}=0, \qquad \av{\hat\epsilon^i \hat\epsilon^j}\equiv\mathbb{C}^{ij},
\end{eqnarray}
here, $\mathbb{C}$ represents the intrinsic statistical fluctuation for the data $\mathbb{C}_{\rm data}$ or the simulations $\mathbb{C}_{\rm mock}$. Moreover, the cross-correlation between the stochastic component of the systematics and the statistical fluctuation of the density field could also be present:
\begin{eqnarray}
   \av{\hat\epsilon_{\rm s}^i \hat\epsilon^j}\equiv \Delta\mathbb{C}_{\rm s}^{ij}.
\end{eqnarray}

In the presence of systematics, the inverse of the covariance receives an additional correction 
\begin{eqnarray}
    \mathbb{\hat C}^{-1} 
    &\rightarrow& (\mathbb{C}+\Delta \mathbb{C}_{\rm s} + \Delta\mathbb{C}_{\rm ss})^{-1} \nonumber\\ 
    &\approx&\mathbb{C}^{-1} - \mathbb{C}^{-1}\left(\Delta\mathbb{C}_{\rm s} + \Delta\mathbb{C}_{\rm ss}\right)\mathbb{C}^{-1}.
\end{eqnarray}
Given Eq.~(\ref{eqn:zeta_3_term}--~\ref{eqn:epsilon_cs}), we find the expectation value of the product of the 4PCF observables is
\begin{eqnarray}
    \av{\hat\zeta^i \hat\zeta^j} = \sum_{a,b \in\{{\rm g, s}\}}\zeta_{\rm a}^i\zeta_{\rm b}^j + \left(\mathbb{C}^{ij} + \Delta\mathbb{C}^{ij}_{\rm s} + \Delta\mathbb{C}^{ij}_{\rm ss}\right).
\end{eqnarray}
As a result, the expectation of the $\chi^2$ defined in Eq.~\eqref{eqn:chi2-auto} under the null assumption, that there is no connected contribution to 4PCF,  becomes:

\begin{eqnarray}
    \av{\chi^2} &=& \av{\hat\zeta^i\, (\mathbb{\hat C}^{-1})_{ij}\, \hat\zeta^j}\\
    &=& (\mathbb{\hat C}^{-1})_{ij}\left(\zeta_{\rm g}^i\zeta_{\rm g}^j + 2\zeta_{\rm s}^i\zeta_{\rm g}^j+ + \zeta_{\rm s}^i\zeta_{\rm s}^j\right) + {\rm Tr}\left[\mathbb{\hat C}^{-1} \mathbb{C}\right]\nonumber,
\end{eqnarray}
where $(\mathbb{\hat C}^{-1})_{ij} \zeta_{\rm s}^i\zeta_{\rm g}^j = (\mathbb{\hat C}^{-1})_{ij} \zeta_{\rm g}^i\zeta_{\rm s}^j$, given that the covaraince matrix is symmetric $\mathbb{C}_{ij} = \mathbb{C}_{ji}$.

The difference between the $\chi^2$ from the data and the mocks is given by
\begin{eqnarray}\label{eqn:chi2_avg_diff_data_mock}
\av{\chi^2_{\rm data}} - \av{\chi^2_{\rm mock}} &=& \av{(\hat\zeta_{\rm data}-\zeta_{\rm g})^i (\mathbb{\hat C}^{-1})_{ij} (\hat\zeta_{\rm data}-\zeta_{\rm g})^j}\nonumber\\
&&-\, \av{(\hat\zeta_{\rm mock}-\zeta_{\rm g})^i (\mathbb{\hat C}^{-1})_{ij} (\hat\zeta_{\rm mock}-\zeta_{\rm g})^j}\nonumber\\
&\equiv& \mathbb{\hat C}^{-1}_{ij} \left(\Delta_{\rm gg}^{ij} + 2\Delta_{\rm gs}^{ij} + \Delta_{\rm ss}^{ij}\right)\nonumber\\
&& +\, {\rm Tr}\left[\mathbb{\hat C}^{-1} (\mathbb{C}_{\rm data} - \mathbb{C}_{\rm mock})\right]
\end{eqnarray}
with the residual difference in cross-correlation products between data and mock catalogs $\Delta^{ij}_{\rm ab}$ defined as

\begin{eqnarray}\label{eqn:Delta_ab}
    \Delta_{\rm ab}^{ij} \equiv \zeta^{{\rm data},i}_{\rm a} \zeta^{{\rm data},j}_{\rm b} - \zeta^{{\rm mock},i}_{\rm a} \zeta^{{\rm mock},j}_{\rm b},
\end{eqnarray}
where ${\rm a} \in \{ {\rm g, s}\}$ and ${\rm b} \in \{ {\rm g, s}\}$ represent gravity-induced  ``${\rm g}$'' or systematic-induced ``${\rm s}$''.

The expectation values of  $\av{\chi^2_{\rm data}}$ and $\av{\chi^2_{\rm mock}}$ are equal only if the mocks perfectly reproduce nonlinear gravitational evolution, all systematics are fully corrected ({\it i.e.}, $\Delta_{\rm ab}^{ij}=0$), and the covariance matrix is estimated without bias. In that case, the $\chi^2$ statistic effectively quantifies the excess signal arising from the gravity-induced connected 4PCF.
In practice, however, these conditions are not always satisfied, and residual systematics may be misinterpreted as gravity-induced nonlinearity, impacting both the signal and the covariance estimation. 


For the DESI-Y1 LRG sample, one major concern is the fiber assignment implementation scheme. Overall, an incomplete sample has larger statistical fluctuations; in particular, we find that different fiber assignment implementations can lead to a difference in covariance of up to 20\% to 30\% (see~\S\ref{sec:incomplete_stat_fluctuations}). 

For these reasons, in the auto-correlation tests we compare the results using four sets of simulations: (1) \ezmock with FFA (2) \abacus with altMTL (reference) (3) \abacus with FFA (4) ``complete'' Abacus. These sets of tests allow us to isolate the effects due to inaccurate estimation of statistical fluctuations. Moreover, they allow us to understand the contribution due to incomplete fiber assignment implementation effects, and potential misinterpretation of gravity-induced non-Gaussianity.

\subsection{Cross-Correlation Test}
\label{sec:cross-corr}
As an additional test of the detection significance, we follow~\cite{hou2022:parity, Krolewski2024:parity}, and divide the full survey into different smaller angular patches, with data vectors labeled $\zeta^{\mu,i},\zeta^{\nu,j},...$. The idea is that if there is an underlying signal, it should show correlation across different patches of the sky. 

In particular,~\cite{Krolewski2024:parity} proposed to further divide $\chi^2$ in Eq.~\eqref{eqn:chi2-auto} into two terms: (1) $\chi^2_{\times}$, which captures spatially-correlated physics such as gravity-induced connected 4PCF (the signal in our context) and (2) $\chi^2_{\Delta}$, which is sensitive to the mismatch in statistical fluctuations between the data and the mocks. 

The signal-sensitive statistic $\chi^2_{\times}$ is defined as
\begin{eqnarray}\label{eqn:chi2-cross}
\chi^2_{\times} = \frac{1}{\calN_\times}\sum_{\mu<\nu} \hat\zeta^{\mu,i}\, (\mathbb{\hat C}^{-1})_{ij} \, \hat\zeta^{\nu,j},  
\end{eqnarray}
with the normalization factor
\begin{eqnarray}
\calN_\times \equiv N_p (N_p-1)/2,
\end{eqnarray}
with $N_p$ being the number of patches, and the normalization given by the combinatorial factor. 

Under the assumption that different patches are uncorrelated, the underlying density field and the systematics are also uncorrelated. The expectation value of the cross correlation between the 4PCF of two different patches reads 
\begin{eqnarray}\label{eqn:cross_two_patch}
    \av{\hat\zeta^{\mu,i} \hat\zeta^{\nu,j}} = \sum_{a,b \in\{{\rm g, s}\}}\zeta_{\rm a}^{\mu,i}\zeta_{\rm b}^{\nu,j} + \mathbb{\hat C}^{\mu, ij} \delta_{\mu\nu}^{\rm K}.
\end{eqnarray}
Here we explicitly include the systematics term $\zeta_{\rm s}$, which can potentially generate a connected parity-even 4PCF at the signal level.

The expectation value of the signal-sensitive statistic $\av{\chi^2_{\times}}$ is
\begin{eqnarray}
    \av{\chi^2_{\times}} &=& \frac{1}{\calN_\times}\sum_{\mu<\nu} \av{\hat\zeta^{\mu, i}\, (\mathbb{\hat C}^{-1})_{ij} \, \hat\zeta^{\nu,j}}\\
    &=& (\mathbb{\hat C}^{-1})_{ij}\Bigg\{\zeta_{g}^i  \zeta_{g}^j + \frac{1}{\calN_\times}\sum_{\mu<\nu} \left(\zeta_{g}^i  \zeta_{\rm s}^{\nu,j} +\zeta_{\rm s}^{\mu,i}  \zeta_{g}^{j} + \zeta_{\rm s}^{\mu,i}  \zeta_{\rm s}^{\nu,j}\right)\Bigg\}.\nonumber
\end{eqnarray}
Here we impose that the signal from gravitational evolution is identical across different patches, while allowing the systematics in the $\mu$-th and $\nu$-th patches to differ in general.
Compared to the parity-odd case in~\cite{Krolewski2024:parity}, the expectation of the cross statistics for simulations does not center around zero, since the simulations do include gravitationally-induced non-Gaussian effects. Moreover, the additional systematics-induced term can potentially contaminate the cross-statistic. Notably, the cross-correlation between the signal and the systematics may lead to a negative correlation.

Comparing the difference between  $\chi^2_{\times}$ for data and mocks, we arrive at the following:
\begin{eqnarray}\label{eqn:chi2x_avg_diff_data_mock}
    &&\av{\chi^2_{\times, {\rm data}}} - \av{\chi^2_{\times, {\rm mock}}} \nonumber \\
    &=& (\mathbb{\hat C}^{-1})_{ij} \Bigg\{\left(\zeta^i_{g, {\rm data}} \zeta^j_{g, {\rm data}} - \zeta^i_{g,  {\rm mock}} \zeta^j_{g, {\rm mock}}\right) \nonumber\\
    &&\,+ \frac{1}{\calN_\times}\sum_{\mu<\nu}\bigg[\left(\zeta^{i}_{g, {\rm data}} \zeta^{\nu,j}_{s, {\rm data}} - \zeta^{i}_{g,  {\rm mock}} \zeta^{\nu,j}_{s, {\rm mock}}\right)\nonumber\\
    &&\,+ \left(\zeta^{\mu,i}_{s, {\rm data}} \zeta^j_{g, {\rm data}} - \zeta^{\mu,i}_{s,  {\rm mock}} \zeta^j_{g, {\rm mock}}\right)\nonumber\\
    &&\,+ \left(\zeta^{\mu,i}_{s, {\rm data}} \zeta^{\nu,j}_{s, {\rm data}} - \zeta^{\mu,i}_{s,  {\rm mock}} \zeta^{\nu,j}_{s, {\rm mock}}\right)\bigg] \Bigg\}\nonumber\\
    &\equiv& \mathbb{\hat C}^{-1}_{ij} \left(\Delta_{\rm gg}^{ij} + \tilde\Delta_{\rm gs}^{ij}
    + \tilde\Delta_{\rm sg}^{ij}
    + \tilde\Delta_{\rm ss}^{ij}\right),
\end{eqnarray}
where we define the patch-wise residual difference in cross-correlation products between data and mock catalogs $\tilde\Delta_{\rm ab}^{ij}$ is defined as
\begin{eqnarray}
\tilde\Delta_{\rm ab}^{ij} \equiv \frac{1}{\calN_\times}\sum_{\mu<\nu} \zeta^{\rm data,\mu}_{\rm a} \zeta^{\rm data, \nu}_{\rm b} - \zeta^{\rm mock,\mu}_{\rm a} \zeta^{\rm mock, \nu}_{\rm b},      
\end{eqnarray}
for $a\in\{\rm g,s\}$ and $b\in\{\rm g,s\}$. Note that when ${\rm a} = {\rm b} = {\rm g}$, the first pure gravity-induced term reduces to the definition in Eq.~\eqref{eqn:Delta_ab}.

If the systematics in the $\mu$-th and the $\nu$-th patches are identical, Eq.~\eqref{eqn:chi2x_avg_diff_data_mock} trivially reduces to the first term of Eq.~\eqref{eqn:chi2_avg_diff_data_mock}. However, when systematics vary between patches, auto- and cross-statistics will yield different results due to their distinct computational approaches. Auto-correlation statistics compute the 4PCF with systematics applied globally across the entire survey volume, while
cross-statistics compute the 4PCF with systematics applied separately to each individual patch before combining results. These two approaches can yield significantly different outcomes when systematic effects vary sufficiently between patches, particularly when the systematic variations are comparable to or larger than the underlying cosmological signal.

Similarly to Eq.~\eqref{eqn:chi2_avg_diff_data_mock}, if the simulations accurately reproduce gravitational nonlinearity and systematics, $\Delta_{\rm ab}^{ij}$ should vanish, and the difference between the two cross statistics should approach zero. In Eq.~\eqref{eqn:chi2x_avg_diff_data_mock} we also see that gravity-induced nonlinearity is not the only cause for the cross-statistic to deviate from zero; if the systematics were not faithfully implemented, they can also potentially lead to a data-mock mismatch. For this purpose, we applied different tests in \S\ref{sec:Result} to quantify the impact of systematics at the signal level.

Note that the parity-even case studied here differs from the test applied to the parity-odd case presented in~\cite{Krolewski2024:parity}, where a discrepancy between the mock and data indicates a parity-violating signal. In the test for the gravitationally induced 4PCF, we are interested in both the agreement between the data and mocks, and in any deviation of the data minus mock distribution from the null distribution, where the latter indicates the detection significance.

As a further consistency check, we now turn to the statistical mismatch-sensitive statistic $\chi^2_{\Delta}$ defined as
\begin{eqnarray}\label{eqn:chi2-Delta}
\chi^2_{\Delta} = \frac{1}{\calN_\Delta} \sum_{\mu<\nu} (\hat\zeta^{\mu}-\hat\zeta^{\nu})^i\, (\mathbb{\hat C}^{-1})_{ij} \, (\hat\zeta^{\mu}-\hat\zeta^{\nu})^j,   
\end{eqnarray}
with $\calN$ being the normalization factor
\begin{eqnarray}\label{eqn:normalization}
    \calN_\Delta \equiv (N_p-1)\sum_{\mu}\frac{V_{\rm eff}}{V_{\rm eff}^{\mu}}.
\end{eqnarray}
Following~\cite{Krolewski2024:parity}, we define the effective volume to be
\begin{eqnarray}\label{eqn:V_eff}
    V_{\rm eff} \equiv \frac{V_{\rm fid} N_{\rm dof}}{{\rm Tr}\left[\mathbb{C}^{-1}_{\rm th}\, \mathbb{\hat C}_{\rm mock}\right]},
\end{eqnarray}
where $V_{\rm fid}$ is the fiducial volume of the survey, obtained by fitting the analytic covariance to the \ezmock. $N_{\rm dof}$ is the number of degrees of freedom for the data vector, $\mathbb{C}_{\rm th}$ is the analytic covariance (see \S\ref{sec:analyt-cov}). If the analytic covariance is a good approximation of the mock covariance $\hat{\mathbb{C}}_{\rm mock}$, $V_{\rm eff} \approx V_{\rm fid}$. However, in our case, due to the difference in the number density in different patches, there could be a significant difference between the two volumes.

The effective volume for each patch is similarly defined as $V_{\rm eff}^{\mu} \equiv {V_{\rm fid} N_{\rm dof}} / {{\rm Tr}\left[\mathbb{C}^{-1}_{\rm th}\, \mathbb{\hat C}^{\mu}\right]}$. Again, if the analytic covariance is a good approximation of the mock covariance, the definition of the effective volume for $\mu$-th patch reduces to a rescaling of the fiducial volume.

The expectation value of the mismatch-sensitive statistic is given as
\begin{eqnarray}\label{eqn:chi2_Delta_avg}
    \av{\chi^2_{\Delta}} &=& \frac{1}{\calN_\Delta}\sum_{\mu<\nu} \av{ \left(\hat\zeta^{\mu}-\hat\zeta^{\nu}\right)^i\, (\mathbb{\hat C}^{-1})_{ij} \, \left(\hat\zeta^{\mu}-\hat\zeta^{\nu}\right)^j}\\
    &=& \frac{N_p-1}{\calN_\Delta}\left\{\sum_\mu {\rm Tr}\left[\mathbb{\hat C}^{-1} \mathbb{\hat C}^{\mu}\right] + \sum_{\mu<\nu} \Delta\zeta_s^{\mu\nu,i} (\mathbb{\hat C}^{-1})_{ij} \Delta\zeta_s^{\mu\nu,j}\right\}\nonumber,
\end{eqnarray}
to obtain the second line, we used the definition of the trace $\mathrm{Tr}\left[\mathbb{\hat C}^{-1} \mathbb{\hat C}^{\mu}\right]=\sum_i \left[\mathbb{\hat C}^{-1} \mathbb{\hat C}^{\mu}\right]_{ii} = \sum_{ij} (\mathbb{\hat C}^{-1})_{ij} \mathbb{\hat C}^{\mu}_{ij}$. We note that the deterministic gravity terms in Eq.~\eqref{eqn:cross_two_patch} cancel between the different patches, while we still keep the potential difference in the systematics-induced bias across different patches $\Delta\zeta_{\rm s}^{\mu\nu,i}\equiv \zeta_{\rm s}^{\mu,i}-\zeta_{\rm s}^{\nu,i}$.

From Eq.~\eqref{eqn:chi2_Delta_avg}, we can see that the normalization coefficient in Eq.~\eqref{eqn:normalization} is defined to recover the number of degrees of freedom of the data vector. We note that the definition assumes that the statistical fluctuations of different patches only differ by an overall rescaling, and the systematics-induced correction is negligible. However, the variation in the number densities and the difference in the fiducial and the effective volume as shown in Table~\ref{tab:statistic} indicate that the statistical fluctuation across different patches may be more than an overall scaling factor. Moreover, there is no guarantee the systematics-induced bias should vanish. Consequently, the center of the $\chi^2_\Delta$ can deviate from the number of degrees of freedom. 

The difference between  $\chi^2_{\Delta}$ for data and mocks is given as follows:
\begin{eqnarray}
    &&\av{\chi^2_{\Delta, {\rm data}}} - \av{\chi^2_{\Delta,{\rm mock}}} \nonumber \\
    &=& \frac{N_p-1}{\calN_\Delta} \sum_\mu {\rm Tr}\left[\mathbb{\hat C}^{-1} (\mathbb{C}^{\mu}_{\rm data} - \mathbb{C}^{\mu}_{\rm mock})\right], 
\end{eqnarray}
which quantifies the mismatch between the statistical fluctuation of the data and mocks.

\subsection{Analytic Covariance Matrix}
\label{sec:analyt-cov}
In this section, we provide a brief review of the analytic covariance matrix derived following the method outlined in~\cite{Hou2022:AnalytCov}. The covariance for the NPCF coefficients in the isotropic basis reads
\begin{eqnarray}\label{eqn:cov_npcf}
&& \mathbb{C}_{\Lambda,\Lambda'}\left(R, R'\right) \nonumber\\
&= & (4 \pi)^{3N/2} \int \frac{s^2 d s}{V} \sum_G \sum_{\Lambda^{\prime \prime}, \calL_G}(-1)^{\sum_i (-\ell_i -\ell_i'+\ell_i'') / 2} \calB_{\calL_G, \Lambda}^{G^{-1}} \nonumber\\
&&\times\calG^{\calL_G \Lambda^{'} \Lambda^{''}} \mathcal{D}_{\Lambda^{''}} \mathcal{C}_0^{\Lambda^{''}} \prod_{i=0}^{N-1} f_{\ell_{G i} \ell_i^{'} \ell''_i}\left(r_{G i}, r_i', s\right)|_{r_{G 0}=r_0^{'}=0}. 
\nonumber \\
\end{eqnarray}
We let $\Lambda \equiv \{\ell_0,\ldots, \ell_{N-1}\}$ and $\Lambda' \equiv \{\ell_0',\ldots, \ell_{N-1}'\}$ denoting the angular momenta associated with the position vectors $\bfR\equiv\{\bfr_0,\ldots,\bfr_{N-1}\}$ and $\bfR'\equiv\{\bfr_0',\ldots,\bfr_{N-1}'\}$. $\bfs$ is the vector denote the separation between the two copies of the $N$-tuplets of the density fields. Its corresponding primary angular momenta are $\Lambda''=\{\ell_0'',\ldots\,\ell_{N-1}''\}$. The $N!$ permutations are given by the notation $G$.  The permutation $G$ can cause the order of the $r_{Gi}$ to differ from the canonical ordering by length.  This accounted for by defining $\calB_{\calL_G, \Lambda}^{G^{-1}}$ through
\begin{align}
\calP_{\calL_G}(\hr_{G1},\hr_{G2}\ldots)=\sum_\Lambda \calB_{\calL_G, \Lambda}^{G^{-1}}\calP_\Lambda(\hr_1,\hr_2,\ldots)
\end{align}
where the $(r_1,r_2,\ldots)$ are ordered by length.

For $N=4$ with only three non-zero elements, the re-ordering operator reduces to a product of Kronecker deltas, $\delta^{\rm K}_{\Lambda_i \calL_{G_i}}$, and is given by
\begin{eqnarray}
\calB_{\calL_G, \Lambda}^{G^{-1}} = (-1)^{\sum_{i=0}^3 \ell_i (1-\calE_{G})/2} \prod_{i=0}^{N-1} \delta^{\rm K}_{\Lambda_i \calL_{G_i}},
\end{eqnarray}
where $\calE_{G}=1$ for even permutations and $\calE_{G}=-1$ for odd permutations.
The generalized Gaunt integral reads~\cite{Cahn202010}
\begin{eqnarray}\label{eqn:generalized_gaunt}
\calG^{\calL_G \Lambda^{'} \Lambda^{''}}= (4 \pi)^{-2}\left[\prod_{i=0}^3 \calD_{\ell_{Gi} \ell_i^{\prime} \ell_i^{\prime \prime}}^{\mathrm{P}} \calC_{000}^{\ell_{Gi} \ell_i^{\prime} \ell_i^{\prime \prime}}\right] \mathcal{Q}^{\calL_G \Lambda^{\prime} \Lambda^{\prime \prime}},
\end{eqnarray}
with 
\begin{eqnarray}\label{eqn:calQ}
\calQ^{\calL_G \Lambda^{'} \Lambda^{''}} = \prod_{i=0}^3 \sum_{m_{Gi} m_i' m_i''} \mathcal{C}_{m_{Gi} m_i^{\prime} m_i^{\prime \prime}}^{\ell_{Gi} \ell_i^{\prime} \ell_i^{\prime \prime}} \mathcal{C}_{\mathrm{M}_G}^{\calL_{G}} \mathcal{C}_{\mathrm{M}^{\prime}}^{\Lambda^{\prime}} \mathcal{C}_{\mathrm{M}^{\prime \prime}}^{\Lambda^{\prime \prime}},
\end{eqnarray}
where
\begin{eqnarray}\label{eqn:calC_l1l2l3_m1m2m3}
\calC^{\ell_1\ell_2\ell_3}_{m_1 m_2 m_3} \equiv \six{\ell_1}{\ell_2}{\ell_3}{m_1}{m_2}{m_3}.
\end{eqnarray}
In the case of $N=4$, we have 
\begin{eqnarray}
\calC^{\Lambda}_{\rm M} &\equiv& \calC^{\ell_0\ell_1(\ell_*)\ell_2\ell_3}_{m_0m_1(m_*)m_2m_3}\\
&=& (-1)^{\sum_{i=0}^3 \ell_i+\ell_*}\calD_{\ell_*}\sum_{m_*}(-1)^{m_*}\calC^{\ell_0 \ell_1 \ell_*}_{m_0 m_1 m_*}  \calC^{\ell_* \ell_2 \ell_3}_{-m_* m_2 m_3}\nonumber,   
\end{eqnarray}
and similarly,
\begin{eqnarray}\label{eqn:calC_Lambda_0}
\calC^{\Lambda''}_{0}&\equiv&\calC^{\ell_0''\ell_1''(\ell_*'')\ell_2''\ell_3''}_{0000}\\
&=& (-1)^{\sum_{i=0}^3 \ell_i''+\ell_*''}\calD_{\ell_*''}\calC^{\ell_0'' \ell_1'' \ell_*''}_{000}  \calC^{\ell_*'' \ell_2'' \ell_3''}_{000}.\nonumber
\end{eqnarray}
and
\begin{eqnarray}\label{eqn:calD_Lambda}
\calD_{\Lambda''}
&=&\calD_{\ell_0''\ell_1''\ell_2''\ell_3''}\nonumber\\
&\equiv& \sqrt{(2\ell_0''+1)(2\ell_1''+1)(2\ell_2''+1)(2\ell_3''+1)},
\end{eqnarray}
Here we choose to fix $\ell_0'=0$ and permute only $\ell$'s. Depending on the position of zeros for the angular momentum $\ell_{G0}$, they will lead to seemingly different structure in the 6-$j$ and 9-$j$ symbols. In Appendix~\ref{sec:generalized_gaunt}, we discuss these three cases and show the unification of the coefficients for these three different cases.

The $f$-function is defined as\
\begin{eqnarray}
   && f_{\ell_{i} \ell_i^{'} \ell_i''}\left(r_{i}, r_i', s\right) \nonumber\\
   &\equiv& \int \frac{k^2 dk}{2\pi^2} j_{\ell_{i}}(k r_{i}) j_{\ell'_{i}}(k r'_{i}) j_{\ell_{i}''}(k s) P(k),
\end{eqnarray}
where $P(k)$ is the power spectrum. In the case of discretised samples, the power spectrum also includes the shot noise: $P(k) \rightarrow P(k) + 1/\bar{n}$, with $\bar{n}$ being the galaxy number density.

Finally, we fit the analytic covariance for its galaxy number density $\bar{n}_{\rm g}$ and the sample volume $V_{\rm fid}$ to the \ezmock-FFA mocks following~\cite{Hou2022:AnalytCov}. The fitted values are summarised in Table~\ref{tab:statistic}. This fitted covariance can partially mitigate effects such as survey geometry and gravitational nonlinearity that are not fully modeled in the analytic approach. However, it does not capture all contributions, including potential systematic effects, which can lead to deviation from the Gaussian distribution assumption (see \S\ref{subsec:nonGaussian_likelihood}).


\subsection{Dimensionality Reduction with Eigendecomposition}
\label{sec:data_compression}
Given the high incompleteness of the DESI DR1 sample, the mock-based covariance offers a more reliable estimate of the covariance matrix. However, due to the high dimensionality of the data vector and the limited number of the available mocks, data vector compression is needed before we proceed with the estimating the detection significance. For this purpose, we use the analytic covariance to compress the data vector. The analytic covariance matrix can be decomposed as
\begin{eqnarray}
    \mathbb{C}_{\rm th} = U \Lambda U^{\rm T},
\end{eqnarray}
with $\Lambda$ being the diagonal matrix of eigenvalues with the diagonal entries $\lambda_i$, and $U$ the orthonormal matrix of eigenvectors, the 4PCF coefficients can be projected onto the eigenbasis of the analytic covariance as
\begin{eqnarray}
    \tilde\zeta = U^{\rm T}\zeta
\end{eqnarray}
Next, we select the components that have the highest signal-to-noise ratio
\begin{eqnarray}
    \text{SNR}_i = \frac{|\tilde\zeta_i|}{\lambda_i^{1/2}},
\end{eqnarray}
where $\tilde\zeta_i$ is the mean of the 4PCF estimated from the simulations. 

This compression method enables us to construct a hybrid covariance matrix by decomposing the 4PCFs measured from simulations (specifically, the 1,000 \ezmock\, with FFA implementation used in this work) into the eigenbasis of the analytic covariance matrix. We select the top $N_{\rm eig}$ eigenmodes with the highest signal-to-noise ratio and use the corresponding rotated data vectors $\tilde\zeta$ to build a covariance matrix in the compressed basis:
\begin{eqnarray}
    \hspace{-6mm}\hat{\mathbb{C}}_{{\rm hybrid}, ij} = \frac{1}{N_{\rm mock}-1}\!\!\sum_{n=1}^{N_{\rm mock}}\!\!\left(\tilde\zeta^{(n)}_i - \av{ \tilde\zeta}_i\right) \left(\tilde\zeta^{(n)}_j - \av{ \tilde\zeta}_j\right).
\end{eqnarray}
Due to the data compression, even though the covariance matrix is estimated from mocks and subject to the condition $N_{\rm sim} \gg N_{\rm dof}$, the reduced size of the data vector allows for a smaller number of simulations to be sufficient.

In \S\ref{sec:Result}, we will compare the results obtained using both the analytic covariance and the hybrid approach, discussing the differences in the detected significance and the implications of assuming a Gaussian likelihood.

\section{Results}
\label{sec:Result}

In this section, we first present the detection significance based on the full auto-correlation statistic $\chi^2$, as defined in Eq.~\eqref{eqn:chi2-auto}. However, as discussed in \S\ref{sec:auto-corr}, the full auto-correlation may be affected by potential data–mock mismatches. To address this, we perform an additional test using the cross statistic $\chi^2_\times$, described in \S\ref{sec:cross-corr}.

\begin{figure*}
    \centering
    \includegraphics[width=0.99\linewidth]{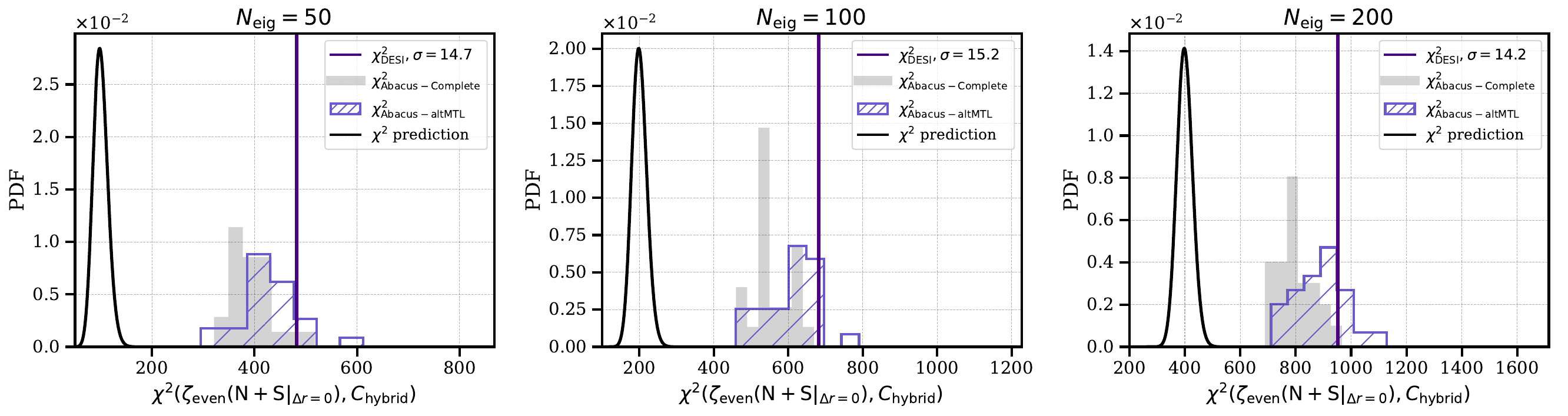}
    \caption{Distribution of the auto-correlation $\chi^2$ for the parity-even connected 4PCF, using the full DESI DR1 LRG sample, combining the NGC and SGC. The panel shows the comparison between simulations and data, as well as the detection significance of the gravitationally-induced connected 4PCF.  
    Each plot corresponds to the number of eigenvalues  used: $N_{\rm eig}=\{50,100, 200\}$. 
    We used the range $20\, \mpch <r<160\,\mpch$. The data vector and the corresponding \ezmock-FFA-based covariance are both compressed according to \S\ref{sec:data_compression}. The vertical lines represent the statistics for the DESI data; the filled gray histograms correspond to the \textsc{Abacus}-Complete; the hashed histogram corresponds to the \textsc{Abacus}-altMTL; the black curves show the \jhc{theoretical predictions of a $\chi^2$ distribution}. We evaluated the significance using both the probability-to-exceed and the Wilson-Hilferty transformation, finding Gaussian-equivalent significances of $14.7\sigma$, $15.2\sigma$, and $14.2\sigma$. In addition, we rescale all the distributions by the ratio of the fiber implementation given by the FFA and the altMTL fiber implementation, which reduces the $\chi^2$ by 15\%--20\%.}
    \label{fig:chi2_hybrid_NS_even_dr0}
\end{figure*}

Fig.~\ref{fig:chi2_hybrid_NS_even_dr0} presents the detection significance for the full DESI DR1 LRG sample, combining both the NGC and SGC regions. Each plot corresponds to a different number of eigenvalues, $N_{\rm eig}=\{50,100, 200\}$. We used the range $20\, \mpch <r<160\,\mpch$  with ten equally spaced radial bins. This range and spacing is used in all the following analyses.  The vertical lines represent the statistics from the DESI data. The predictions assuming the absence of any connected 4PFC are shown as black curves. For further consistency checks, we show the distributions from the \abacus\ simulations: one using the altMTL implementation (hatched histogram), and the other from the Complete mocks without systematic effects (filled histogram), which allow us to isolate the impact of pure nonlinear gravitational evolution. We find that the two distributions from simulations are consistent with each other and with the DESI data, indicating that systematic effects do not artificially enhance the detection significance or lead to a misinterpretation of the gravity-induced connected parity-even 4PCF. 

Since the observed $\chi^2_{\rm obs}$ is much higher than the number of degrees of freedom for the data vector, $\chi^2_{\rm obs} \gg \chi^2_{\rm dof}$, a probability-to-exceed (PTE)-based significance estimation will run into numerical instability. We therefore assess significance using the Wilson–Hilferty transformation~\cite{Wilson1931:WH} as a statistical approximation that maps a chi-squared distributed variable into an approximately normal distributed variable. Suppose $X\sim \chi^2_{\rm obs}$, the Wilson–Hilferty transformation reads

\begin{eqnarray}\label{eqn:sigma_WH}
    \sigma_{\rm WH} \equiv \frac{(X/N_{\rm dof})^{1/3}- (1-2/(9N_{\rm dof}))}{\sqrt{2/(9N_{\rm dof})}}.
\end{eqnarray}
For DESI DR1 LRG sample, we obtain Gaussian-equivalent values of $14.7\sigma$, $15.2\sigma$, and $14.2\sigma$ as shown in Fig.~\ref{fig:chi2_hybrid_NS_even_dr0}. We note that the Wilson–Hilferty transformation could yield a more conservative estimation of the significance. Alternatively, one can obtain a more accurate estimate of the tail probability using a saddle point approximation. The two approaches yield similar answers.

\begin{figure*}
    \centering
    \includegraphics[width=0.99\linewidth]{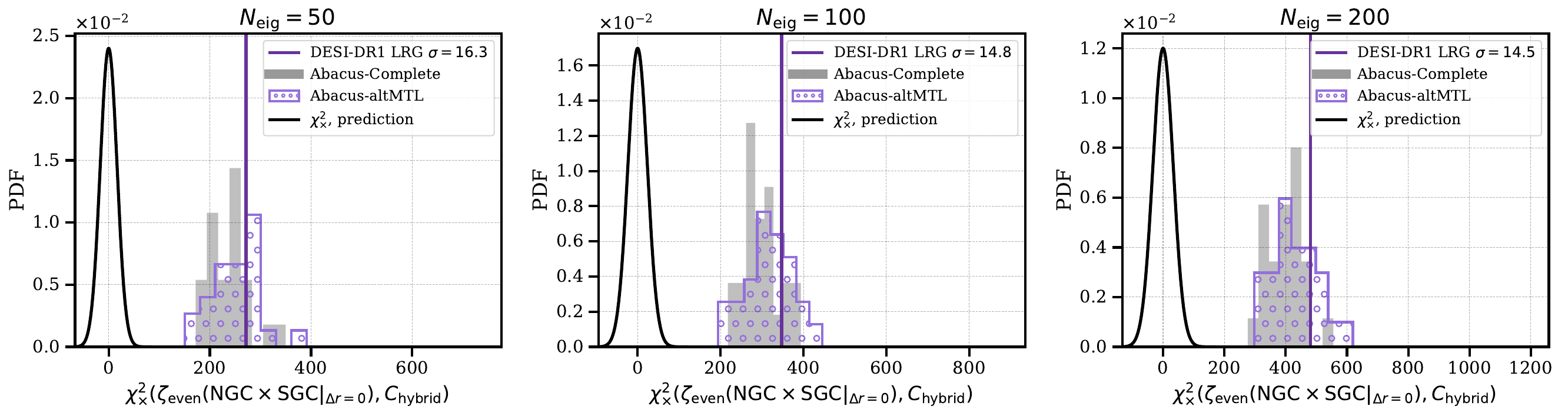}
    \caption{Distribution of the cross-correlation $\chi^2_\times$ for the parity-even connected 4PCF, using the full DESI DR1 LRG sample, combining the NGC and SGC. \jhc{The black curves show the theoretical predictions for the cross statistic under the null hypothesis, assuming a Gaussian distribution centered at zero with a variance given by Eq.~\eqref{eqn:var_chi2x}}.
    The rest of the labeling is identical to that in the preceding figure.}
    \label{fig:chi2_x_hybrid_NS_even_dr0}
\end{figure*}

Since we applied the data compression scheme as in \S\ref{sec:data_compression}, the covariance matrix is essentially based on the \ezmock-FFA. As we will see in\S\ref{sec:incomplete_stat_fluctuations}, the FFA fiber assignment implementation leads to an underestimation of the covariance compared to the altMTL implementation. We therefore rescale all distributions by the ratio between the FFA and altMTL samples, leading to a 15\%–20\% reduction in $\chi^2$. Additionally, we also apply the Hartlap factor $M\equiv (N_{\rm m}-N_{\rm b} -2)/(N_{\rm m} -1)$ to de-bias the inverse of the sample covariance~\cite{Hartlap2007:hartlap}.

In addition to the auto-correlation, we perform a cross-correlation test as discussed in \S\ref{sec:cross-corr}. 
Fig.~\ref{fig:chi2_x_hybrid_NS_even_dr0} shows the detection significance of the cross-correlation statistic $\chi^2_\times$ for the DESI DR1 LRG sample between NGC and SGC. Each plot shows results for a different number of eigenvalues, $N_{\rm eig} = \{50, 100, 200\}$.
Both the data vector and the associated \ezmock-FFA-based covariance matrix are compressed as described in \S\ref{sec:data_compression}. Vertical lines indicate the $\chi^2_\times$ values obtained from the DESI data. The filled gray histograms represent the distribution from the \textsc{Abacus}-Complete mocks, while the histogram marked with circles corresponds to the \textsc{Abacus}-altMTL mocks. The theoretically predicted distribution of the signal-sensitive cross statistic is shown as the black curve. Under the null hypothesis, {\it i.e.}, the signal term in Eq.~\eqref{eqn:cross_two_patch} vanishes $\av{\hat\zeta^\mu\hat\zeta^\nu}_{\zeta^{(c)}=0} \rightarrow \delta^{\rm K}_{\mu\nu} \mathbb{C}^{\mu}$, leading to $\av{\chi^2_{\times, \mathrm{null}}} = 0$. As a result, the black curve of Fig.~\ref{fig:chi2_x_hybrid_NS_even_dr0} is centered around zero.

Next, to further quantify the significance of the signal-sensitive cross statistic, we evaluate its variance, which is defined by 
\begin{eqnarray}\label{eqn:def_variance}
\text{Var}\left(\chi^2_{\times, \mathrm{null}}\right) = \av{\left(\chi^2_{\times, \mathrm{null}}\right)^2} - \av{\chi^2_{\times, \mathrm{null}}}^2,
\end{eqnarray}
where the first term is
\begin{eqnarray}
\av{\left(\chi^2_{\times, \mathrm{null}}\right)^2} 
&=& \frac{1}{\calN_\times^2} \sum_{\substack{\mu< \nu\\ \sigma<\rho}} \av{\hat{\zeta}^\mu_i (\hat{\mathbb{C}}^{-1})_{ij}
\hat\zeta^\nu_j \hat{\zeta}^\rho_k (\hat{\mathbb{C}}^{-1})_{kl}\hat\zeta^\sigma_l}.
\end{eqnarray}
This is to be evaluated in the Gaussian approximation where the four-point expectation is a sum of products of two-point expectations.  The second term of Eq. (\ref{eqn:def_variance}) cancels the contribution where $\mu=\nu$ and $\sigma=\rho$.  The remaining four-point expectation is

\begin{eqnarray}
\av{\hat{\zeta}^\mu_i\hat\zeta^\nu_j\hat{\zeta}^\rho_k\hat\zeta^\sigma_l}_{\substack{\mu<\nu\\ \sigma<\rho}} =  \delta^{\rm K}_{\mu\sigma}\hat{\mathbb{C}}^\mu_{il} \delta^{\rm K}_{\nu\rho}\hat{\mathbb{C}}^\nu_{jk}.     
\end{eqnarray}
Inserting this, we have for the variance of the signal-sensitive cross statistic under the null hypothesis (also see~\cite{Krolewski2024:parity})
\begin{eqnarray}\label{eqn:var_chi2x}
    \text{Var}(\chi^2_{\times, {\rm null}}) &=& \frac{1}{\calN_\times^2} \sum_{\mu< \nu} \text{Tr} \left[\mathbb{C}^{-1}\, \mathbb{\hat C}^{\mu}\, \mathbb{C}^{-1}\, \mathbb{\hat C}^{\nu}\right] \nonumber\\
    &\approx& \frac{N_{\rm dof}}{\calN_\times^2}  \sum_{\mu< \nu}  \frac{V^2_{\rm fid}}{V^\mu_{\rm eff} V^\nu_{\rm eff}},
\end{eqnarray}
where $V_{\rm eff}^{\mu}$ is the effective volume for each patch, defined in Eq.~\eqref{eqn:V_eff}. Here we choose to normalize all the patches with respect to the full NGC, and thus $V_{\rm fid}$ corresponds to the volume of the full NGC. Here we approximate the distribution of $\chi^2_\times$ as a Gaussian distribution, with Eq.~\eqref{eqn:var_chi2x}, the width of the black curve is $\sigma(\chi^2_{\times})|_{N_{\rm eig}=50}=16.6$, $\sigma(\chi^2_{\times})|_{N_{\rm eig}=100}=23.5$, and $\sigma(\chi^2_{\times})|_{N_{\rm eig}=200}=33.2$.
To test the estimation of the variance for the cross statistic in the absence of a real signal, we also analyze the parity-odd modes in Appendix~\ref{sec:odd_4pcf}. The good agreement between the theoretical prediction and the mock distribution in Appendix Fig.~\ref{fig:chi2_x_hybrid_NS_odd_dr0} demonstrates the validity of the estimation.

As in Fig.~\ref{fig:chi2_hybrid_NS_even_dr0},
a fiber correction factor is also applied to  account for the use of the FFA configuration in the \ezmock\ implementation. Comparing the significance found in Fig.~\ref{fig:chi2_hybrid_NS_even_dr0} and Fig.~\ref{fig:chi2_x_hybrid_NS_even_dr0}, we found good agreement between both the autocorrelation and the cross-correlation.

\begin{figure*}
    \centering
    \includegraphics[width=0.99\linewidth]{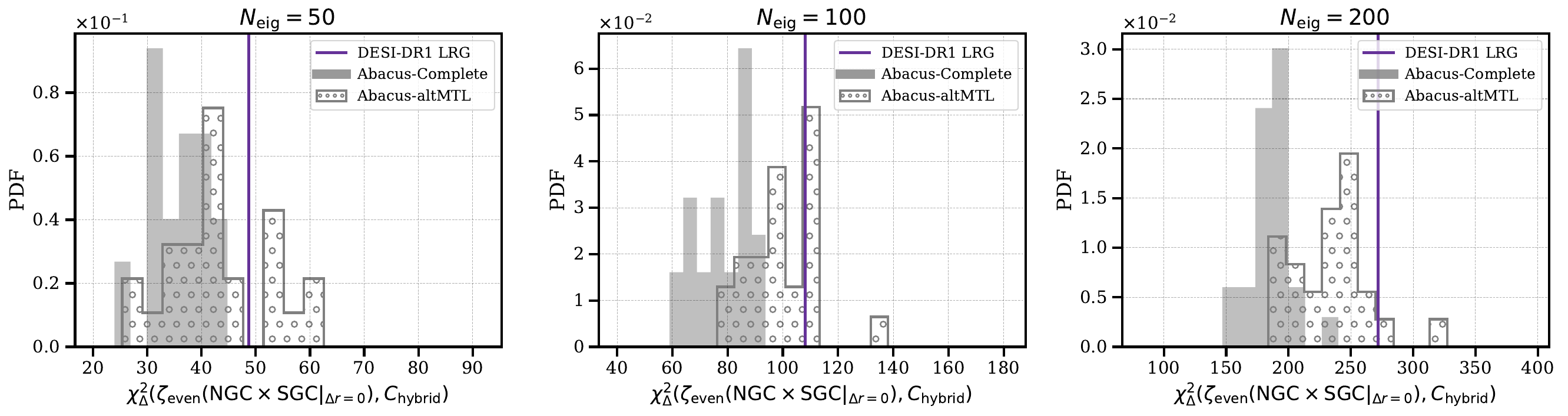}
    \caption{Distribution of the $\chi^2_\Delta$ statistic for the parity-even connected 4PCF, using the full DESI DR1 LRG sample, combining the NGC and SGC. Each plot corresponds to the number of eigenvalues $N_{\rm eig}=\{50,100, 200\}$. 
    The data vector and the corresponding \ezmock-FFA-based covariance are both compressed according to \S\ref{sec:data_compression}. The vertical lines represent the statistics for the DESI data; the filled gray histograms correspond to the \textsc{Abacus}-Complete mocks. The histogram marked by circles corresponds to the \textsc{Abacus}-altMTL mocks.}
    \label{fig:chi2_null_hybrid_NS_even_dr0}
\end{figure*}

As a further consistency check, we test null cross-correlation $\chi^2_{\times, {\rm null}}$ for the DESI DR1 LRG sample between NGC and SGC as shown in Fig.~\ref{fig:chi2_null_hybrid_NS_even_dr0}. This statistic is targeted at distinguishing the data-mock mismatch. As before, each plot here corresponds to the number of eigenvalues $N_{\rm eig}=\{50,100, 200\}$. 
The data vector and the corresponding \ezmock-FFA-based covariance are both compressed according to \S\ref{sec:data_compression}. The vertical lines represent the statistics for the DESI data, the filled gray histograms correspond to the \textsc{Abacus}-Complete mocks, the histogram marked with circles corresponds to the \abacus-altMTL mocks. Here, we also apply the fiber correction factor. We find good agreement between the \abacus-altMTL mocks and data across the three numbers of eigenvalues, indicating that the statistical fluctuations estimated from the mocks are consistent with the intrinsic fluctuations in the data. Moreover, the distribution also centers around the number of of degrees of freedom, as predicted in Eq.~\eqref{eqn:chi2_Delta_avg}.

\begin{figure}
    \centering
    \includegraphics[width=0.9\linewidth]{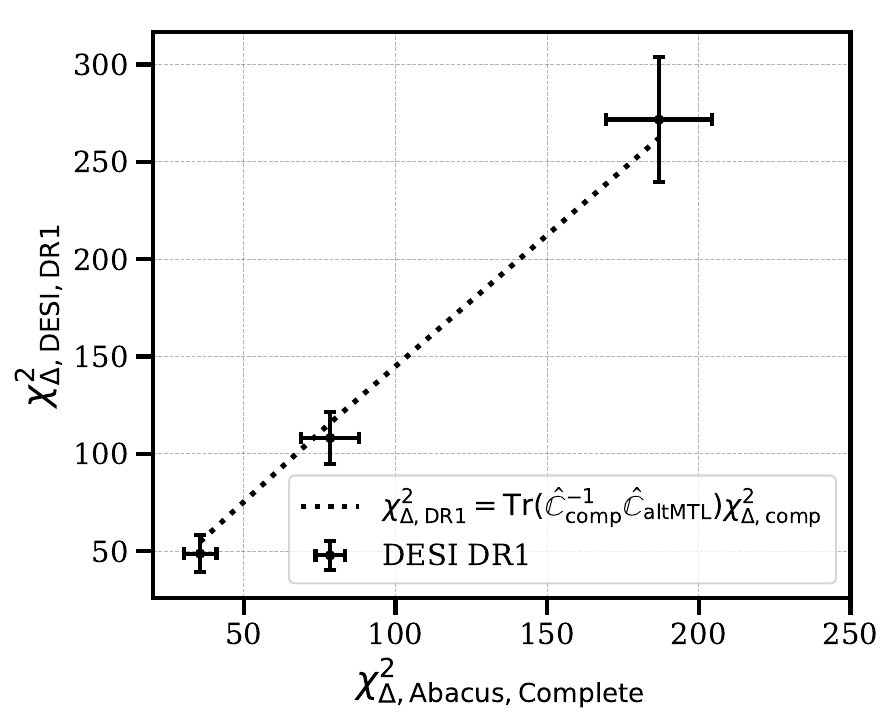}
    \caption{$\chi^2_{\Delta}$ for data and \abacus-Complete mocks. The relation between the two data sets can well be described using the ratio between the statistical fluctuation due to the fiber assignment effects. This demonstrates that the second term in Eq.~\eqref{eqn:chi2_Delta_avg} is negligible.}
    \label{fig:chi2x_Delta_abcs_offset}
\end{figure}

At the same time, we find deviation between the data and the \abacus-Complete mocks and the deviation becomes larger as we increase the number of eigenvalues. The difference could potentially arise from two effects: larger statistical fluctuation due to incompleteness (see~\S\ref{sec:incomplete_stat_fluctuations}) and systematics-induced bias (see second term of Eq.~\ref{eqn:chi2_Delta_avg}). To distinguish these two effects, we further compare the observed $\chi^2_{\Delta}$ for the DESI DR1 LRG sample and the \abacus-Complete mocks as shown in Fig.~\ref{fig:chi2x_Delta_abcs_offset}, with the error bars derived from the \abacus mocks. We find that the slope of the dashed line is well described by the ratio between the fiber assignment using altMTL-assigned \abacus mocks and the one without fiber assignment, demonstrating that the systematic bias in Eq.~\eqref{eqn:chi2_Delta_avg} does not contribute significantly and has a
subdominant impact on the signal.

We highlight that despite the good agreement between the data and \abacus-altMTL, the center value of the distribution is higher than the number of degrees of freedom ({\it c.f.} Eq.~\ref{eqn:chi2_Delta_avg}), which may be a result of the large variation in incompleteness across different patches.

\begin{figure}
    \centering
    \includegraphics[width=0.95\linewidth]{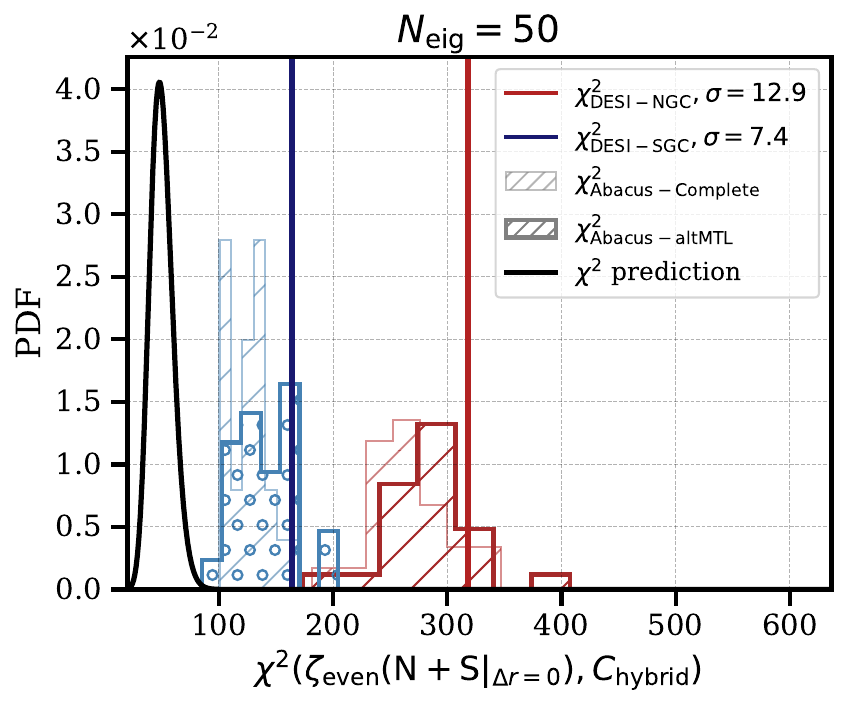}
    \caption{Distribution of the auto-correlation $\chi^2$ for the parity-even connected 4PCF. Here we show the DESI DR1 LRG sample for the NGC (red) and SGC (blue) separately with $N_{\rm eig}=50$. 
    The vertical lines represent the statistics for the DESI data. The thinner-lined histograms correspond to the \abacus-Complete mock. The thicker-lined histograms are for the \abacus-altMTL. The black curve shows the theoretical prediction in the absence of a connected 4PCF.}
    \label{fig:chi2_hybrid_N_S_even_dr0}
\end{figure}

Fig.~\ref{fig:chi2_hybrid_N_S_even_dr0} displays the detection significance for the DESI DR1 LRG sample, shown separately for the NGC and SGC, for $N_{\rm eig}=50$. 
The vertical lines indicate the statistics from the DESI data; the line-hatched histograms correspond to the \abacus-Complete mocks, the circle-hatched histograms to the \abacus-altMTL mocks, and the black curves represent the theoretical predictions. The distributions for both mocks are very close to each other, indicating that systematics do not artificially enhance the detection significance. While here we only display the result for $N_{\rm eig}=50$, we also find very similar results for $N_{\rm eig}=100$ and $N_{\rm eig}=200$.

Using Eq.~\eqref{eqn:sigma_WH}, we find the significances for the NGC and SGC to be $12.9\,\sigma$ and $ 7.4\sigma$. Approximating the two distributions to be independent Gaussian detections, the combined significance is approximately $\sqrt{\sigma_{\rm NGC}^2+\sigma_{\rm SGC}^2}\approx14.9\sigma$, which is in good agreement with the significance obtained from analysing the combined sample directly.

\section{Discussion}
\label{sec:discussion}

In this section, we discuss alternative configurations for the detection significance, including the scale dependence, impact of individual patches as listed in Table~\ref{tab:statistic}, assumption of the Gaussianity of the likelihood, and the impact of sample incompleteness.

\subsection{Scale Dependence}
\label{sec:scale_dependence}

\begin{figure*}
    \centering
    \includegraphics[width=0.99\linewidth]{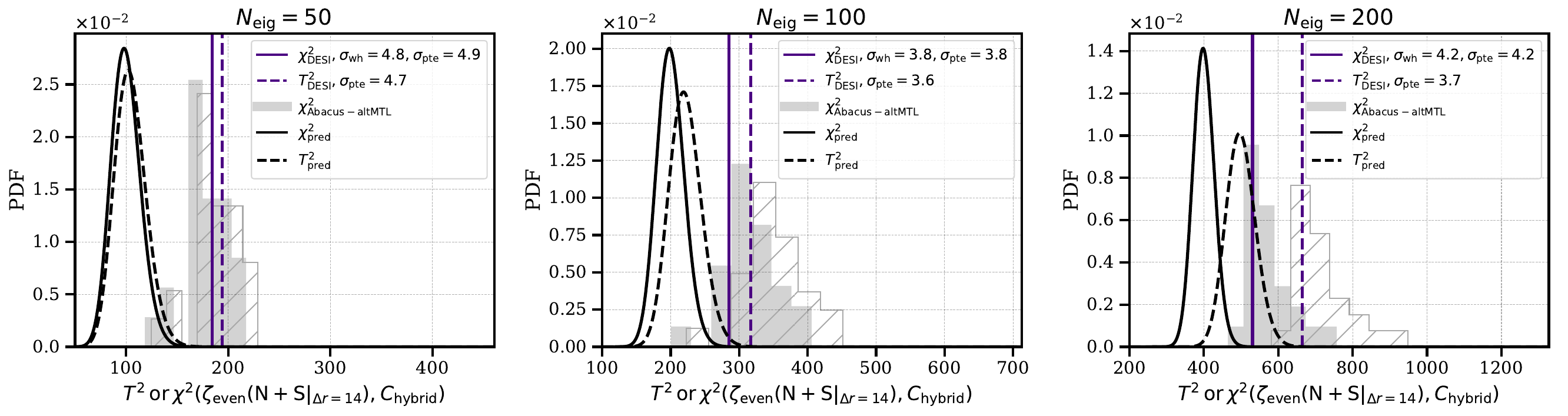}
    \caption{Distribution of the auto-correlation $\chi^2$ and $T^2$ for the parity-even connected 4PCF, using the full DESI DR1 LRG sample, combining the NGC and SGC. Each plot corresponds to the number of eigenvalues $N_{\rm eig}=\{50,100, 200\}$. We 
    applied additional radial cuts $\Delta r=14\,\mpch$, to ensure that used galaxies do not fall in adjacent radial bins. We quantified the detection significance using two statistical distributions: the $T^2$ (dashed) and the $\chi^2$ (solid). The vertical lines represent the statistics for the DESI data, the histograms correspond to the \ezmock, and the black curves show the theoretical predictions. For the $\chi^2$ distribution, we evaluated the significance using both the probability-to-exceed and the Wilson-Hilferty transformation, finding Gaussian-equivalent significances of $4.8\sigma$, $3.8\sigma$, and $4.2\sigma$. The probability-to-exceed method yielded nearly identical values. For $T^2$, we used only the probability-to-exceed approach and obtained detection significances of $4.7\sigma$, $3.6\sigma$, and $3.7\sigma$.}
    \label{fig:chi2_T2_hybrid_NS_even_dr14}
\end{figure*}

Fig.~\ref{fig:chi2_T2_hybrid_NS_even_dr14} shows the detection significance for the full DESI DR1 LRG sample combining NGC and SGC. Each plot corresponds to the number of eigenvalues $N_{\rm eig}=\{50,100, 200\}$. 
We applied additional radial cuts $\Delta r=14\,\mpch$ to enforce a minimum separation between the radial bins, to ensure that used galaxies do not fall into adjacent radial bins.

Here we additionally quantified the detection significance using two statistical distributions: the $T^2$ (dashed) and the $\chi^2$ (solid). The $T^2$ distribution is used because the covariance matrix is estimated from simulations, which modifies the likelihood to follow a $T^2$ form~\cite{Sellentin201602}. This distribution has been adopted in previous analyses~\cite{hou2022:parity,Philcox2021boss4pcf}, with an explicit expression given, {\it e.g.}, in equation (20) of~\cite{hou2022:parity}.

The vertical lines represent the statistics for the DESI data, the histograms correspond to the \ezmock, and the black curves show the theoretical predictions for the null hypothesis. Both the data and the simulations show strong evidence for the presence of a connected four-point correlation. Due to the reduced detection, we can assess the significance using both the Wilson-Hilferty transformation and the probability-to-exceed by computing the area as the fraction of the distribution that lies beyond the detection significance observed in the data.

For the $\chi^2$ distribution, we evaluated the significance using both methods, finding Gaussian-equivalent significances of $4.8\sigma$, $3.8\sigma$, and $4.2\sigma$ using the Wilson-Hilferty transformation. The probability-to-exceed method yielded nearly identical values. For $T^2$, we used only the probability-to-exceed approach and obtained detection significances of $4.7\sigma$, $3.6\sigma$, and $3.7\sigma$.
Compared to the full range of scales in Fig.~\ref{fig:chi2_hybrid_NS_even_dr0}, we find a reduction in the detection, as expected, since the gravity-induced signal should be manifest mostly on small scales. 

\subsection{NGC-1, NGC-2, SGC-3}
\begin{figure*}
    \centering
    \begin{subfigure}[b]{0.3\textwidth}
    \includegraphics[width=0.99\linewidth]{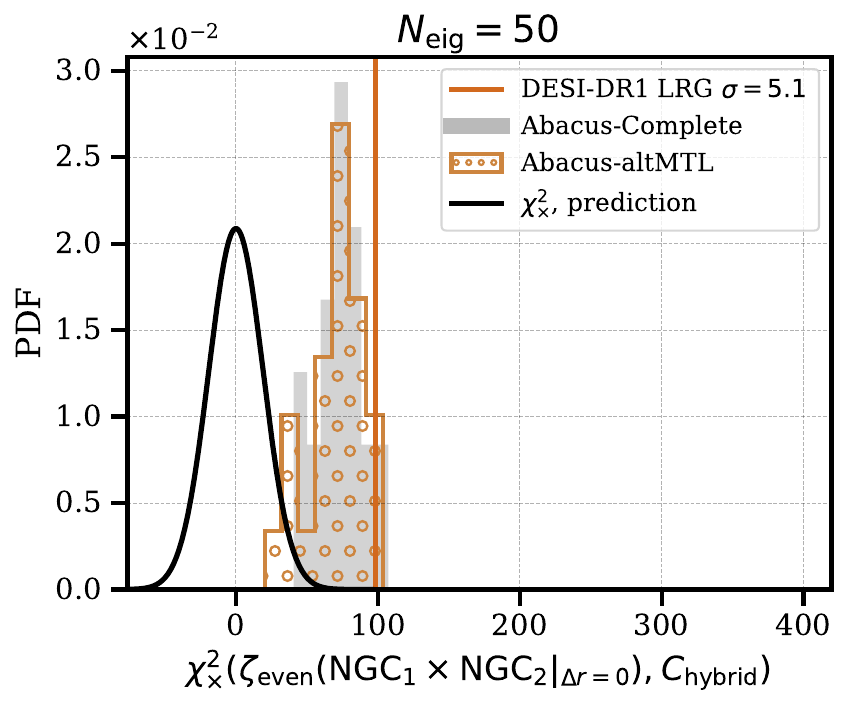}
    \end{subfigure}
    \begin{subfigure}[b]{0.3\textwidth}
    \includegraphics[width=0.99\linewidth]{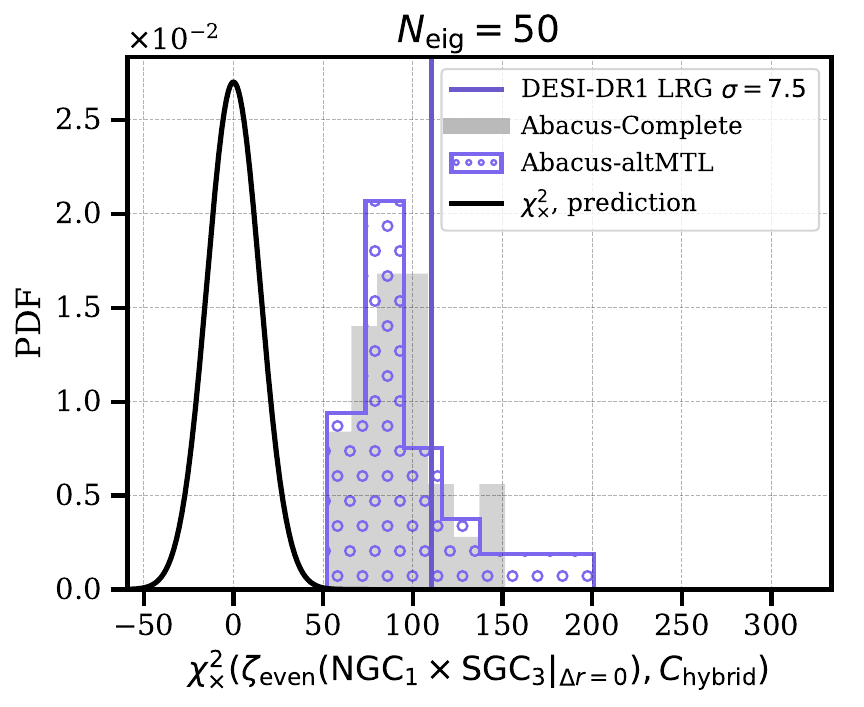}
    \end{subfigure}
    \begin{subfigure}[b]{0.3\textwidth}
    \includegraphics[width=0.99\linewidth]{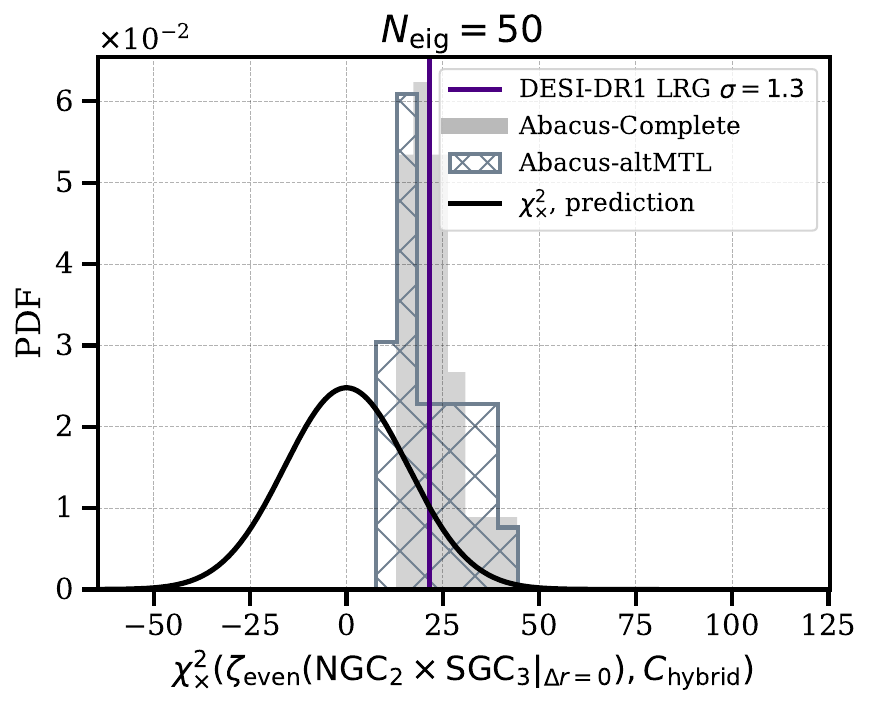}
    \end{subfigure}
    \caption{Distribution of the cross-correlation $\chi^2_\times$ for the parity-even connected 4PCF of the three patches: NGC-1, NGC-2, SGC-3. Here we only show $N_{\rm eig}=50$, given that the significance varies only  marginally for different numbers of eigenvalues. 
      The filled histograms correspond to the \abacus-Complete mocks. The hatched histograms and histograms marked with circles are for the \abacus-altMTL mocks. The vertical line corresponds to the measurement from the data and the black curves show the theoretical predictions for the $\chi^2_\times$ distribution with null hypothesis. For all three plots, we also apply the fiber correction factor.}
    \label{fig:chi2_x_hybrid_N1N2S3_even_dr0}
\end{figure*}

In this section, we examine the significance individually for NGC-1, NGC-2, and SGC-3. As before, we focus on the case with $N_{\rm eig} = 100$, since the results do not show notable variation with the number of eigenmodes. Fig.~\ref{fig:chi2_x_hybrid_N1N2S3_even_dr0} shows the significance for the cross-correlation of the three combinations. From left to right, we have NGC-1$\times$NGC-2, NGC-1$\times$SGC-3, and NGC-2$\times$NGC-3, with significance of $3.6\sigma$, $8.5\sigma$, and $1.6\sigma$, respectively. In addition, we also find good agreement between the data measurement and the mock distribution, with or without fiber assignment. Due to the additional geometry cuts for each patch, the volume for each patch is smaller than the full volume, and we account for it when computing the variance of $\chi^2_\Delta$ ({\it c.f.}, Eq.~\ref{eqn:var_chi2x}). Approximating the three distributions to be independent Gaussian detections, the combined significance is $\sigma_{\rm tot}\approx\sqrt{\sigma_{\rm ngc-1}^2+\sigma_{\rm ngc-2}^2+\sigma_{\rm sgc-3}^2}\approx 9.4$, which is less compared to the fiducial result due to the angular and radial cuts.  

\subsection{Testing Gaussianity of the Likelihood}
\label{subsec:nonGaussian_likelihood}

When analyzing high-dimensional data vectors, one needs to be careful about the assumption of Gaussianity in the likelihood function. Many statistical tests applied to the data rely on this assumption, under which the information can be effectively compressed into a single $\chi^2$ statistic and interpreted in terms of a detection significance, expressed in units of standard deviations relative to a Gaussian distribution. However, when the likelihood deviates significantly from Gaussianity, this interpretation may no longer be valid, and one must be cautious in drawing conclusions based on the $\chi^2$ (or $T^2$)-derived significance.

\begin{figure*}
    \centering
    \begin{subfigure}{0.45\textwidth}
        \centering
        \includegraphics[width=\linewidth]{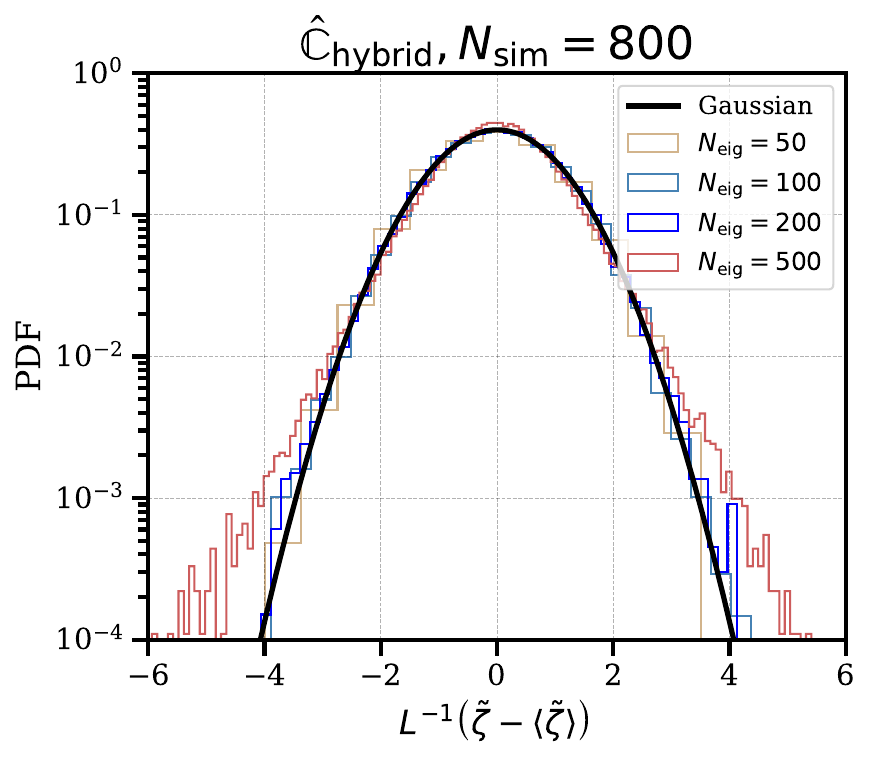}
    \end{subfigure}
    \begin{subfigure}{0.45\textwidth}
        \centering
        \includegraphics[width=\linewidth]{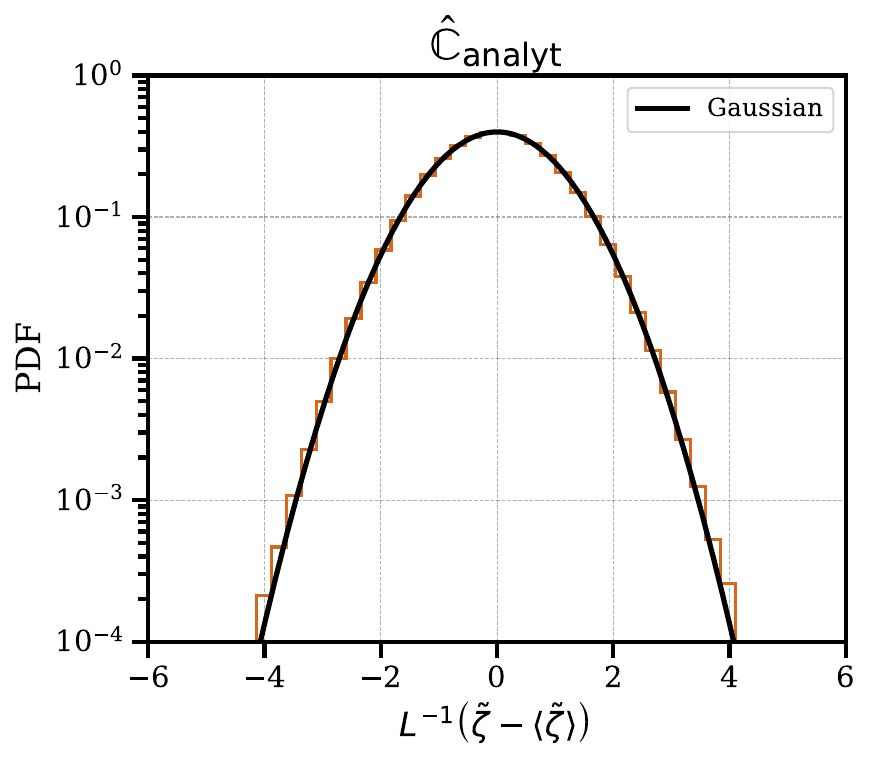}
    \end{subfigure}    
    \caption{The histograms show the distribution of the normalized components of the data vector in mode space, sampled across all eigenmodes and simulations, for different choices of $N_{\rm eig}$ (indicated by color). The black dashed line denotes the standard normal distribution $\mathcal{N}(0,1)$. In the left, this mode space normalization is performed using the inverse Cholesky decomposition $L^{-1}$ of the projected covariance matrix $\hat{\mathbb{C}}_{\rm ezmock} = LL^\top$, estimated from a subset of the 1,000 \ezmock\ realizations. In both panels, the mock set is split into two: one subset is used to estimate the covariance matrix, and the other to generate the test data vectors. The left panel uses 800 mocks for covariance estimation and 200 for testing. When the condition $N_{\rm cov} \gg N_{\rm eig}$ is not well satisfied, the normalized data vector shows deviations from Gaussianity, indicating that the Gaussian likelihood assumption breaks down in this regime. In the right panel, we use the analytic covariance matrix and, despite its various simplified assumptions, find good agreement with the Gaussian likelihood.}
    \label{fig:NG_PDF_Ncov}
\end{figure*}

Following~\cite{Philcox2021boss4pcf}, we assess the validity of the Gaussian likelihood assumption by examining the distribution of Cholesky-normalized data vectors in mode space~\cite{Press2007:numerical}. Fig.~\ref{fig:NG_PDF_Ncov} presents the distribution of the normalized components of the data vector in mode space, sampled across all eigenmodes and simulations, for different choices of degrees of freedom. The black dashed line denotes the standard normal distribution $\mathcal{N}(0,\sigma_{\rm std})$. In the left and middle plots, this mode space normalization is performed using the inverse Cholesky decomposition $L^{-1}$ of the projected covariance matrix $\hat{\mathbb{C}}_{\rm ezmock} = LL^\top$, estimated from a subset of the 1,000 \ezmock\ realizations. In both plots, the mock set is split into two: one subset is used to estimate the covariance matrix, and the other to generate the test data vectors. The left plot uses 800 mocks for covariance estimation and 200 for testing. We find that when the condition $N_{\rm cov} \gg N_{\rm eig}$ is not well satisfied, the normalized data vector shows deviations from Gaussianity, indicating that the Gaussian likelihood assumption breaks down in this regime. 

In the right panel, we test the Gaussian likelihood assumption with the analytic covariance matrix. Although the analytic approach relies on simplified treatments of survey geometry and sample incompleteness, it nevertheless shows good consistency with a Gaussian likelihood. We further discuss the detection significance in Appendix~\ref{sec:significance_analyt_cov}.

\subsection{Impact of Incompleteness due to Statistical Fluctuations}
\label{sec:incomplete_stat_fluctuations}
Due to the limited number of fibers and the large sample, the DR1 sample exhibits low completeness and is therefore sensitive to the details of fiber assignment implementation.
As discussed in \S\ref{sec:desi_dr1}, two fiber assignment implementation schemes were applied to the DESI mocks: the fast fiber assignment (FFA) and the alternate MTL (altMTL).
In the following, we briefly describe these two implementations and examine their impact on the covariance matrices.

We start by briefly describing the altMTL implementation. Each target sample has its own ``Merged Target Ledger'' (MTL)~\cite{Myers2023:DESItargetSelection}, which tracks the observation history of the corresponding targets, including targets' metadata, observation status, redshift information, priority, and etc.~\cite{Lasker2025:DESIFiberAssign}.
In addition to the priority, a subpriority field is generated to assign small additional priority which \textsc{fiberassign} uses to reproducibly resolve collisions.
This is a random number in the range 0 to 1. With the altMTL approach, the default is the inverse completeness weighting $w_{\rm comp}=(f_{\rm TLID} f_{\rm tile})^{-1}$, with $f_{\rm TLID}$ the number of targets competing for a single fiber and $f_{\rm tile}$ is a tile-based completeness that quantifies the fraction of targets that are considered to have been ``observed'' within a group of overlapping tiles ({\it c.f.} Eq.~\ref{eqn:weights}).

The FFA is an emulator that learns the mapping between an input and a \textsc{fiberassign}-processed galaxy catalog, specifically predicting the probability of a target being assigned a fiber based on its properties and tiling information. This mapping function, $\mathcal{F}$, depends on two key variables: the close-friend count, $n_{\rm cfc}$, and the number of tiles, $n_{\rm tile}$. To compute $n_{\rm cfc}$, a Friend-of-Friend (FOF) algorithm is applied to identify galaxies that are closely separated in angular space. The learning algorithm is optimized by comparing to the reference algorithm, in this case the output of the \textsc{fiberassign} code. With the FFA approach, while the default weight is also the inverse completeness weighting, it does not decompose $f_{\rm TLID}$ and $f_{\rm tile}$, but rather learns the average effect of a mixed priory targets such as QSO and LRG as a function of number of overlapping tile $n_{\rm tile}$. 

\begin{figure*}
    \centering
    \includegraphics[width=0.8\linewidth]{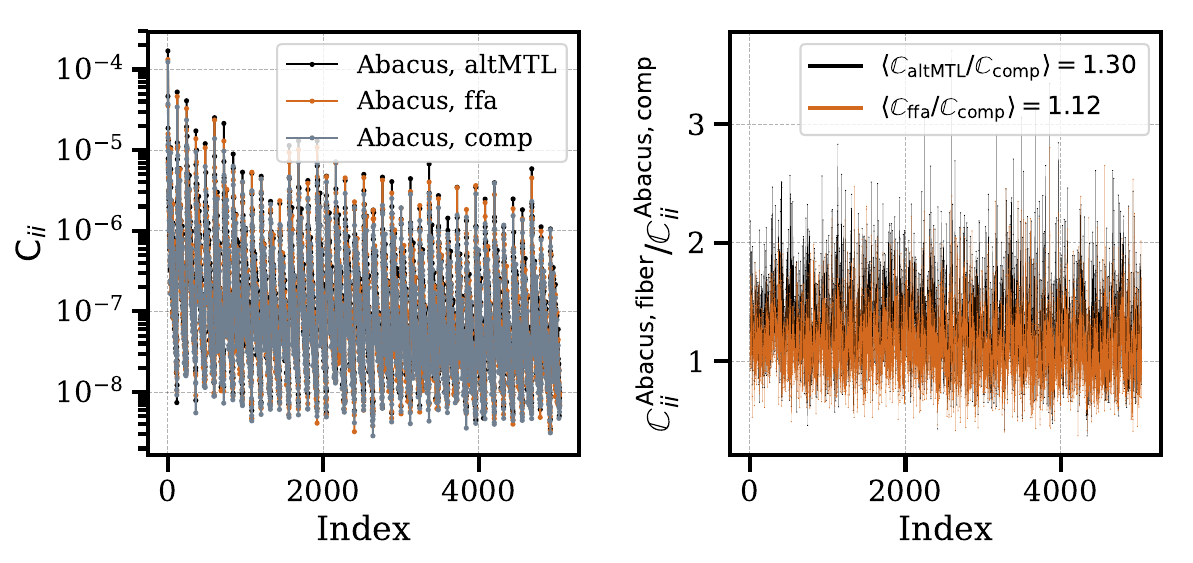}
    \caption{The plot shows the statistical fluctuations in the Abacus simulations under different fiber assignment implementations. {\it Left}: Diagonal elements of the covariance matrix, with different colors indicating altMTL (black), FFA (orange), and a set of mocks without fiber assignment (complete, gray).  {\it Right}: Ratio between the fiber-implemented mocks and the complete mocks. We find increases of 12\% and 30\%  when fiber assignment is included through altMTL and FFA. The covariance is measured from the mocks for the NGC; we find a comparable ratio in the SGC.}
    \label{fig:ratio_cov_NGC_even_abacus}
    
\end{figure*}

Fig.~\ref{fig:ratio_cov_NGC_even_abacus} shows the impact of different fiber assignment schemes on the statistical fluctuations. Left panel shows the diagonal elements of the covariance matrix from the Abacus altMTL (black), Abacus FFA (orange), and Abacus complete (gray). Right panel shows the ratio between the diagonal elements of
Abacus simulations with different fiber assignment implementations and the complete sample. We find the average ratio between the Abacus altMTL to the complete sample (black) to be $\av{\mathbb{C}_{\rm altMTL} / \mathbb{C}_{\rm comp} }\approx 1.12$, whereas the average ratio between the Abacus FFA to the complete sample (orange) to be $\av{\mathbb{C}_{\rm FFA} / \mathbb{C}_{\rm comp} }\approx 1.30$.

A reduced covariance in the FFA mocks was also observed in~\cite{DESI2024:SampleDefinition,ForeroSanchez2025:DESIcovariance}.
Several factors may contribute to the differences in statistical fluctuations as outlined in~\cite{Bianchi2024:DESIfiber}:
altMTL is a more realistic implementation, whereas FFA is based on an approximation. FFA exhibits a less pronounced collision window (i.e., the angular scales affected by the fiber collision effect) and therefore yields fewer zero-probability pairs compared to the altMTL mocks. For a fixed number density, this implies that the FFA-based mocks exhibit smaller density fluctuations, consistent with the observed reduction in covariance.
We leave a detailed quantitative investigation to future work. However, since we use the FFA-based \ezmock\ for covariance estimation, we account for the differences between the two fiber assignment schemes by applying an additional correction factor to our significance, as discussed in \S\ref{sec:Result}.

\section{Summary}\label{sec:summary}
In this paper we presented the measurement of the nonlinear gravity-induced parity-even connected 4PCF on the DESI DR1 LRG sample and quantified its detection significance. We found a 14.7$\sigma$ significance for our fiducial configuration using an auto-correlation test with the number of eigenvalues $N_{\rm eig}=50$. Increasing the number of eigenmodes by a factor of two or four changes the significance by  $\pm 3\%$, indicating robustness to the choice of eigenmode cutoff in our data compression method.

In addition, we tested various analysis setups, including scale dependence by imposing additional radial bin cuts, partitioning the sky into different patches and testing various combinations, and performing both auto- and cross-correlation tests. Excluding smaller scales reduces the significance to $4.8\sigma$, as expected, since small scales are more sensitive to nonlinear gravitational evolution. In contrast, combining sky patches in different ways or using the cross-correlation test yields results consistent with the fiducial configuration. In particular, the two cross statistics allow us to isolate the impact of general systematics by comparing the data to the \abacus-complete mocks. We find that systematics have 
subdominant impact on the signal.

To compute the significance, we compared the $\chi^2$ and $T^2$ distributions and evaluated the probability using both the probability-to-exceed method and the Wilson–Hilferty transformation, finding only a marginal impact on the resulting detection significance.

In this paper, we also briefly discuss the Gaussian likelihood assumption and assess its validity by examining the distribution of Cholesky-normalized data vectors in mode space. We find that the Gaussian assumption holds well for simulation-based hybrid covariance estimates, provided that the number of simulations significantly exceeds the number of degrees of freedom by at least a factor of two $N_{\rm sim} > 2 N_{\rm dof}$. For the analytic covariance, despite the simplifying assumptions involved, we still find good agreement with the Gaussian likelihood. However, the auto-correlation test is more affected by the data–mock mismatch when using the analytic covariance, as discussed in Appendix~\ref{sec:significance_analyt_cov}. We therefore present the hybrid method as our main result.

\begin{itemize}
    \item {\bf Comparison to BOSS}: We re-evaluate the detection significance for the BOSS CMASS sample using the Wilson–Hilferty transformation (Eq.~\ref{eqn:sigma_WH}) and obtain $10.5\ \sigma, 10.4\ \sigma$ and $10.5\ \sigma$ using, 10, 50 and 100 eigenvalues.  We note that when the observed $\chi^2_{\rm obs}$ lies in the extreme tail of the distribution, the probability-to-exceed method can become numerically unstable due to machine precision limitations. The BOSS CMASS sample contains 777,202 galaxies with a completeness exceeding 95\%. Scaling the significance with the square root of the galaxy number yields an expected significance of approximately 17$\sigma$ for DESI. After accounting for the impact of fiber assignment reduces this estimate, bringing it into agreement with our findings in this paper.
    \item {\bf Fiber assignment and sample incompleteness}: We find that the fiber assignment implementation  has a significant impact on the covariance (see \S\ref{sec:incomplete_stat_fluctuations}). This is likely due to the highly incomplete DESI DR1 sample. To account for this, we introduce a correction factor in our significance estimation. Nevertheless, when using the hybrid covariance, we find consistent results between the fiber-assigned mocks and the corresponding complete mocks without fiber assignment. The impact of fiber assignment is expected to be reduced in future data releases with increased sample completeness
    \item {\bf Imaging systematics}: \jhc{For DESI LRG, major imaging systematics were identified in the DECam region, including $E(B-V)$ maps, $r$-band depth, and stellar density~\cite{DESI2024:SampleDefinition}. Since SGC uses entirely DECam photometry while NGC uses a mix of DECam and BASS/MzLS photometry, we tested cross-correlations between NGC and SGC regions to assess whether these DECam-specific systematics bias clustering measurements. The consistency between NGC-SGC cross-correlations and their respective auto-correlations demonstrates that these imaging systematics do not substantially contaminate the clustering signal used for cosmological inference. In addition, the angular integral constraint~\cite{Chaussidon2025:PNG} was also considered but showed marginal impact on LRG measurements when using linear regression weights~\cite{DESI2024:SampleDefinition}. Since this effect primarily manifests on large scales, we do not explicitly test it here, though it should be verified for other tracers when quantifying the connected 4PCF.}
    \item {\bf Parity-odd 4PCF measurements}: Unlike the parity-even 4PCF, the parity-odd signal has a much lower signal-to-noise ratio and is largely noise-dominated, making its detection particularly sensitive to the covariance estimate. However, the DESI DR1 LRG sample suffers from high incompleteness, and the available mocks have limitations: the \ezmock\ underestimates the covariance, while the \abacus\ mocks do not have sufficient volume. As a result, despite performing both auto-correlation and cross-correlation tests, we cannot draw a definitive conclusion regarding the parity-odd result in the DR1 LRG sample. We plan to revisit this analysis with future data releases. Consistency tests based on the parity-odd 4PCF analysis are presented in Appendix~\ref{sec:odd_4pcf}. 
\end{itemize}

\section*{Acknowledgments}
We thank Eiichiro Komatsu, Drew Jamieson, Ariel Sánchez and Sánchez's group for useful discussions. We also thank Zachary Brown, Dragon Huterer, Alex Krolewski, Lado Samushia, and Benjamin Weaver for providing helpful comments on the manuscript. JH thanks~{\hypersetup{urlcolor=black}\href{https://github.com/Moctobers/Acknowledgement/blob/main/fox_in_office.jpg}{Jue Fox}} for his office support. JH has received funding from the European Union’s Horizon 2020 research and innovation program under the Marie Sk\l{}odowska-Curie grant agreement No. 101025187. 
We acknowledge UFIT Research Computing for providing computational resources and support that have contributed to the research results reported in this publication. 
The work of RNC was supported in part by the Director, Office of Science, Office of High Energy Physics, of the U.S. Department of Energy under contract No. DE-AC02-05CH11231.

This material is based upon work supported by the U.S. Department of Energy (DOE), Office of Science, Office of High-Energy Physics, under Contract No. DE–AC02–05CH11231, and by the National Energy Research Scientific Computing Center, a DOE Office of Science User Facility under the same contract. Additional support for DESI was provided by the U.S. National Science Foundation (NSF), Division of Astronomical Sciences under Contract No. AST-0950945 to the NSF’s National Optical-Infrared Astronomy Research Laboratory; the Science and Technology Facilities Council of the United Kingdom; the Gordon and Betty Moore Foundation; the Heising-Simons Foundation; the French Alternative Energies and Atomic Energy Commission (CEA); the National Council of Humanities, Science and Technology of Mexico (CONAHCYT); the Ministry of Science, Innovation and Universities of Spain (MICIU/AEI/10.13039/501100011033), and by the DESI Member Institutions: \url{https://www.desi.lbl.gov/collaborating-institutions}. Any opinions, findings, and conclusions or recommendations expressed in this material are those of the author(s) and do not necessarily reflect the views of the U. S. National Science Foundation, the U. S. Department of Energy, or any of the listed funding agencies.

The authors are honored to be permitted to conduct scientific research on I'oligam Du'ag (Kitt Peak), a mountain with particular significance to the Tohono O’odham Nation.

\section{Data Availability}
Data from the plots in this paper is available on Zenodo as part of DESI’s Data Management Plan~\footnote{\url{https://doi.org/10.5281/zenodo.16729072}}.

\appendix

\section{Parity-odd 4PCF}
\label{sec:odd_4pcf}

In this section, we consider the parity-odd 4PCF, focusing on the consistency between the statistical fluctuations observed in simulations and those predicted by theoretical expectations. We present the consistency check for both the auto-correlation and the cross-correlation tests. 

\begin{figure*}
    \centering
    \includegraphics[width=0.95\linewidth]{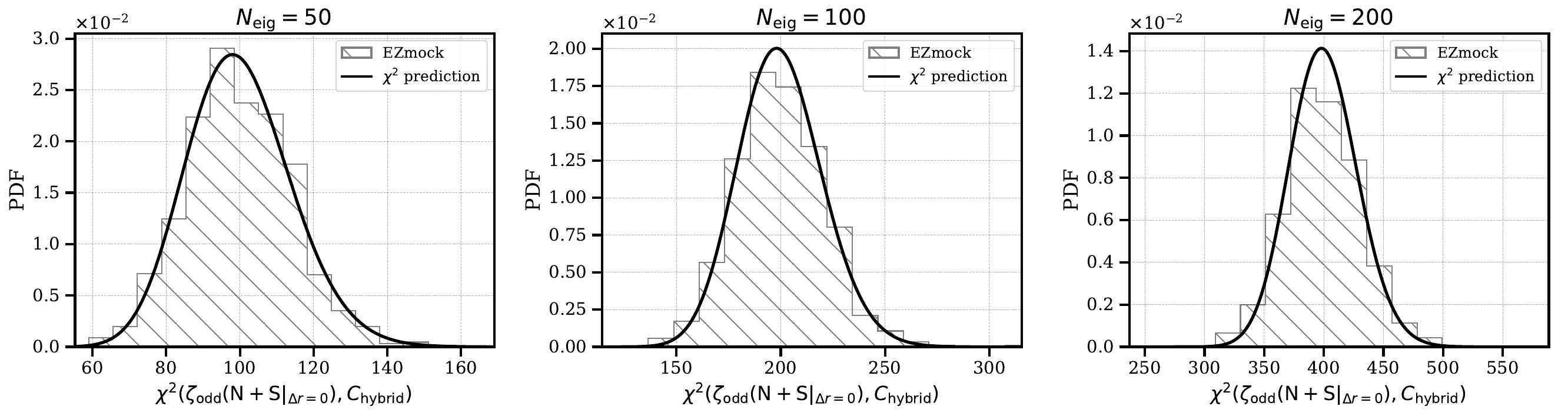}
    \caption{Distribution of $\chi^2$ for the parity-odd 4PCF combining the NGC and SGC using \ezmock\ only. We do not display measurements from DESI Y1 LRG. The fiber correction reduces the data $\chi^2$ by $10\%$ in the odd sector.  The black curve is the theoretical prediction in the absence of parity-violation. For different choices of eigenmodes, we find the \ezmock\ distributions agree well with the theoretical $\chi^2$ distribution.}
    \label{fig:chi2_hybrid_NS_odd_dr0}
\end{figure*}

Fig.~\ref{fig:chi2_hybrid_NS_odd_dr0} shows the auto-correlation for the parity-odd 4PCF combining the NGC and SGC. We do not show the measurements from the DESI data in the parity-odd case. We find that the correction due to the fiber assignment reduces the data $\chi^2$ by $10\%$. 
For the auto correlation test, we use only the \ezmock-FFA, which has sufficient volume to reproduce the statistical fluctuation due to cosmic variation. We do not include the \abacus\ results here, as the box side length of the \abacus\ mocks is only one-third that of the \ezmock\ simulations. This leads to increased statistical fluctuations and artificially high $\chi^2$ values. The black curve is the theoretical prediction for a $\chi^2$ distribution.  The \ezmock\ $\chi^2$ distribution is well described by the theoretical prediction.

\begin{figure*}
    \centering
    \includegraphics[width=0.95\linewidth]{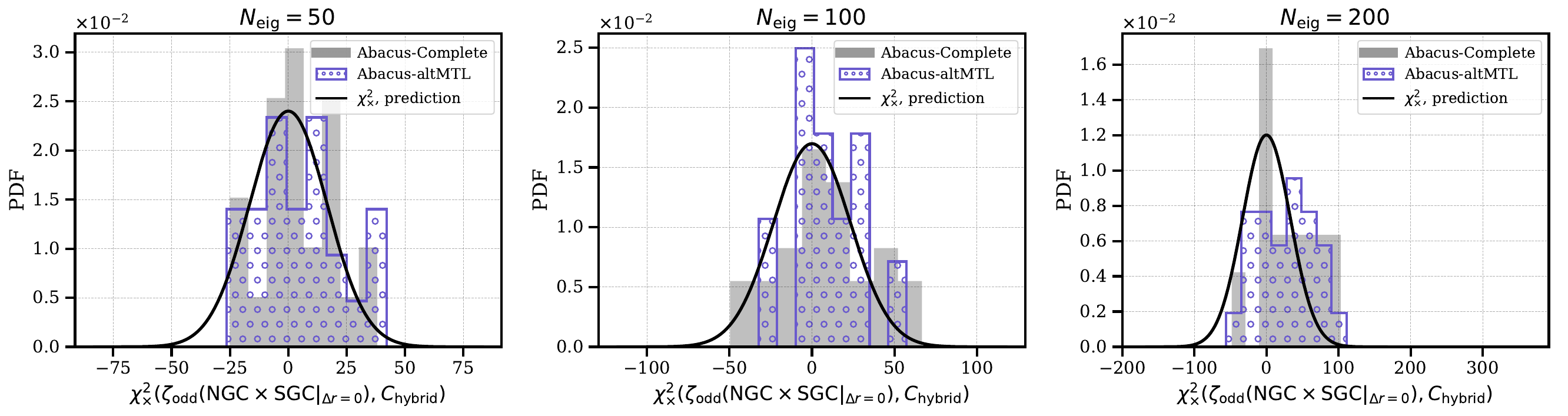}
    \caption{Cross correlation $\chi^2_\times$ of the parity-odd 4PCF between NGC and SGC using \ezmock\ .  We do not display measurements from DESI Y1 LRG.
    The \ezmock-FFA-based covariance is compressed according to \S\ref{sec:data_compression}. The filled gray histograms correspond to the \textsc{Abacus}-Complete mocks; the histogram marked with circles corresponds to the \textsc{Abacus}-altMTL mocks. The black curves show the predictions for the cross statistic under the null hypothesis, assuming a zero-mean Gaussian distribution with variance given by Eq.~\eqref{eqn:var_chi2x}.}
    \label{fig:chi2_x_hybrid_NS_odd_dr0}
\end{figure*}

Fig.~\ref{fig:chi2_x_hybrid_NS_odd_dr0} shows the cross correlation $\chi^2_\times$ of the parity-odd 4PCF between NGC and SGC using \ezmock\ only. 
Since statistical fluctuations do not contribute to the cross-correlation, the smaller volume of the \abacus\ mocks does not have  adirect consequence on the inferred $\chi^2_\times$, and we include all available mocks for this test.

The \ezmock-FFA-based covariance is compressed according to \S\ref{sec:data_compression}. 
The filled gray histograms correspond to the \textsc{Abacus}-Complete mocks, and the histogram marked with circles corresponds to the \textsc{Abacus}-altMTL mocks. The black curves show the predictions for the cross statistic under the null hypothesis, assuming a Gaussian distribution with a zero mean with variance given in Eq.~\eqref{eqn:var_chi2x}. A fiber correction factor is applied to account for the fact that the covariance is calibrated with respect to \ezmock-FFA. 

We note that the good agreement between the theoretical prediction (black curve) and the mock distribution demonstrates that Eq.~\eqref{eqn:var_chi2x} provides a reliable estimate of the cross-statistic $\chi^2_\times$ in the absence of a real signal.

\section{Detection Significance with Analytic Covariance}
\label{sec:significance_analyt_cov}
\begin{figure*}
    \centering
    \begin{subfigure}[b]{0.33\textwidth}
    \includegraphics[width=0.99\linewidth]{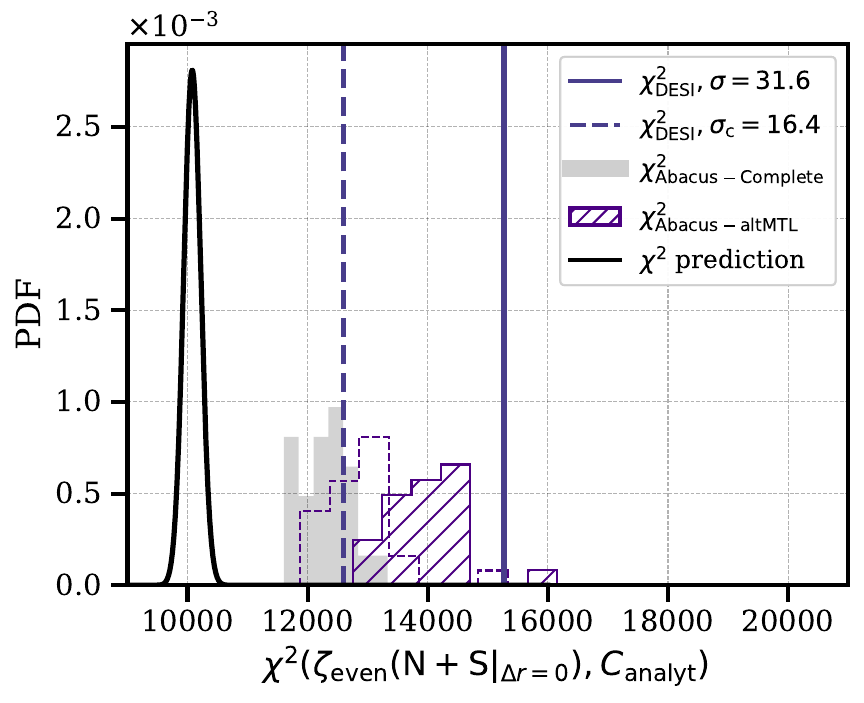}
    \end{subfigure}
    \begin{subfigure}[b]{0.31\textwidth}
    \includegraphics[width=0.99\linewidth]{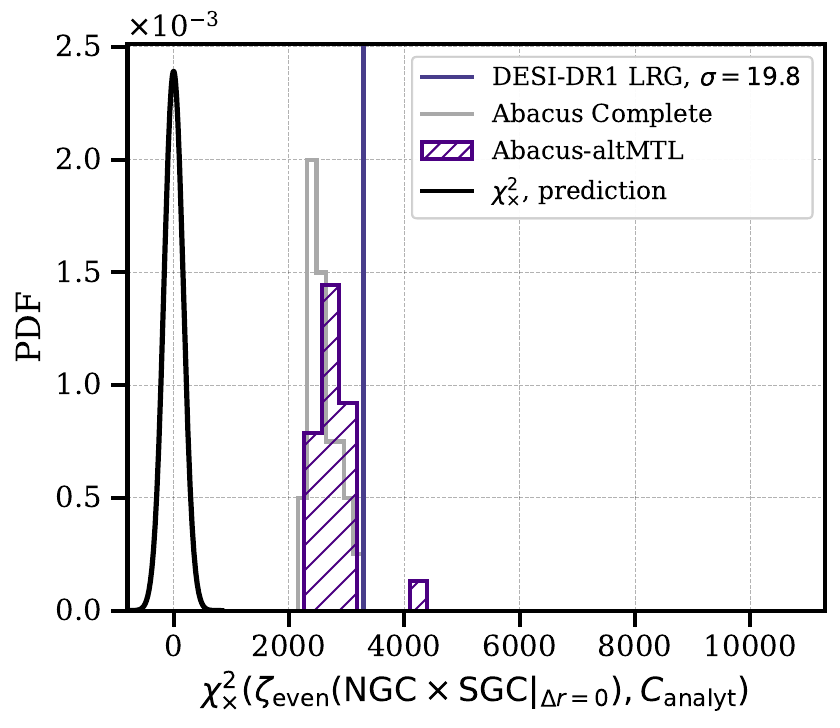}
    \end{subfigure}
    \begin{subfigure}[b]{0.31\textwidth}
    \includegraphics[width=0.99\linewidth]{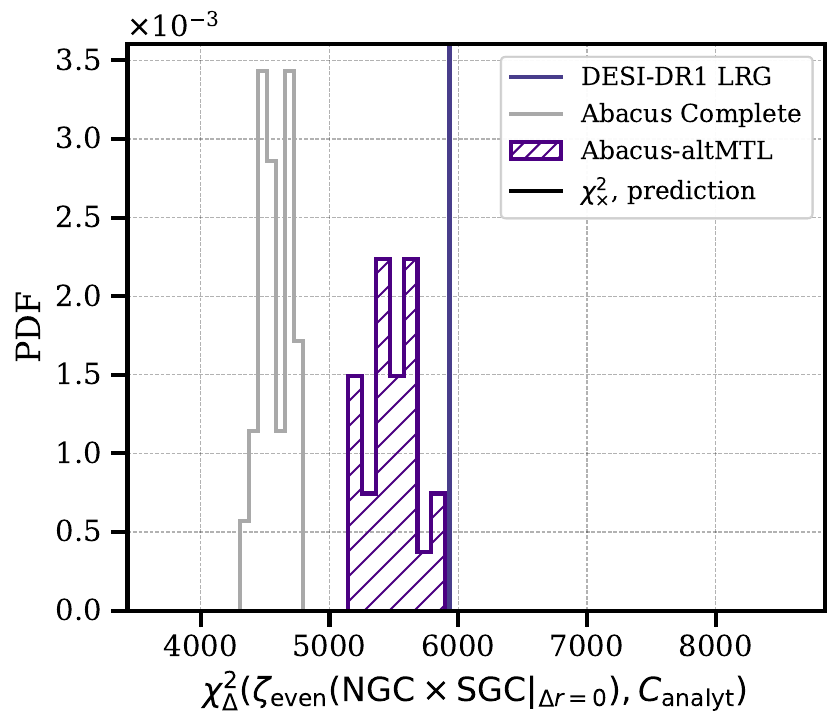}
    \end{subfigure}
    \caption{Auto-statistic (left), cross-statistic $\chi^2_\times$ (middle), and  cross-statistic $\chi^2_\Delta$ (right) of the parity-even 4PCF measured from the full DESI DR1 LRG sample in both the NGC and SGC, using the analytic covariance matrix. The vertical lines represent the statistics for the DESI data.  The histograms correspond to the \abacus mocks and the black curves show the theoretical predictions in the absence of a connected 4PCF.}
    \label{fig:chi2_analyt_NS_even_dr0}
\end{figure*}

In this section, we show the results using the analytic covariance. In Fig.~\ref{fig:chi2_analyt_NS_even_dr0}, we show the detection significance using the auto-statistic (left), cross-statistic $\chi^2_\times$ (middle), and  cross-statistic $\chi^2_\Delta$ (right) of the parity-even 4PCF measured from the full DESI DR1 LRG sample in both the NGC and SGC, using the analytic covariance matrix. In all three panels, the vertical lines are the measurements from the DESI DR1 LRG sample, while the solid and hatched histograms correspond to the \abacus-altMTL.

As discussed in \S\ref{sec:auto-corr}, the auto-correlation can suffer from data–mock mismatch, leading to an overestimation of the detection significance. In the left panel, the uncorrected value shown by the solid vertical line corresponds to a $31.6\sigma$ detection significance, which is 60\% higher compared to the cross-statistic in the middle panel. At the same time, there is a discrepancy between the data and the mock distribution, as well as a disagreement between the two \abacus\ mocks. To account for this, and following the procedure in \S\ref{sec:cross-corr}, we apply $\chi^2_\Delta$ to correct for the mismatch. Specifically, we first compute the difference between the two \abacus\ mocks,
$\delta\chi^2_{\Delta,1} \equiv \chi^2_{\Delta,{\rm altmtl}} - \chi^2_{\Delta,{\rm comp}}$,
which quantifies the systematic-induced fluctuations. We then compute the difference between the data and the \abacus-altMTL,
$\delta\chi^2_{\Delta,2} \equiv \chi^2_{\Delta,{\rm iron}} - \langle \chi^2_{\Delta,{\rm comp}} \rangle$,
which characterizes the residual fluctuation.

In the left panel, we apply $\delta\chi^2_{\Delta,1}$ to the \abacus-altMTL distribution (dashed histogram) and $\delta\chi^2_{\Delta,2}$ to the data (dashed vertical line). After these corrections, the mock distributions show good agreement with each other, and the resulting detection significance is $16.4\sigma$, consistent with the value obtained using the compressed method.

\section{Generalized Gaunt Integral}
\label{sec:generalized_gaunt}
In this section, we take a different approach to derive the analytic covariance, without explicitly introducing the ``H'' permutation as done in~\cite{Hou2022:AnalytCov}. From Eq.~\eqref{eqn:cov_npcf}, we first recognize that the prefactor of interest is
\begin{eqnarray}
\calF^{\calL_G \Lambda^{'} \Lambda^{''}} 
&=& \calE_{G}\calG^{\calL_G \Lambda^{'} \Lambda^{''}} \calD_{\Lambda''}\calC^{\Lambda''}_{0}\\
&=& (4\pi)^{-2}\,\calE_{G}\prod_{i=0}^3 \calD_{\ell_{Gi}\ell_i'\ell_ii''} \calC^{\ell_{Gi}\ell_i'\ell_i''}_{000} \nonumber\\  &&\, \times\,\calQ^{\calL_G \Lambda^{'} \Lambda^{''}}\calD_{\Lambda''}\calC^{\Lambda''}_{0}\nonumber,
\end{eqnarray}
where $\calE_{G}$ denotes a permutation of $\{\ell_1,\ell_2,\ell_3\}$, with $\calE_{G}=1$ for an even permutation of the three indices, and with $\calE_{G}=-1$ for an odd permutation. The generalized Gaunt integral $\calG^{\calL_G \Lambda^{'} \Lambda^{''}}$ is defined in Eq.~\eqref{eqn:generalized_gaunt}. The coefficients $\calD^{\ell_{Gi}\ell_i'\ell_i''}_{000}$ and $\calC^{\ell_{Gi}\ell_i'\ell_i''}_{000}$ are defined in Eq.~(\ref{eqn:calD_l1l2l3} -- \ref{eqn:calC_l1l2l3}), respectively, while $\calD_{\Lambda''}$ and $\calC^{\Lambda''}_0$ are defined in Eqs.~(\ref{eqn:calD_Lambda} -- \ref{eqn:calC_Lambda_0}). As mentioned in \S\ref{sec:analyt-cov}, we fix $\ell_1'=0$, such that $\ell_*'\rightarrow \ell_1'$ and $m_*'\rightarrow m_1'$. Given the  definition of $\calQ^{\calL_G \Lambda^{'} \Lambda^{''}}$ in Eq.~\eqref{eqn:calQ}, we have
\begin{eqnarray}
\calQ^{\calL_G \Lambda^{'} \Lambda^{''}} &=& \sum_{\rm M M' M''} (-1)^{\ell_{G0}+\ell_1'-m_{G0}-m_1'} \calD^{-1}_{\ell_{G0}\ell_1'} \delta^{\rm K}_{\ell_{G0}\ell_0''} \delta^{\rm K}_{m_{G0},-m_0''} \nonumber\\
&&\,\times \sum_{\substack{\ell_* \ell_*'', m_* m_*''}} (-1)^{\sum_{i=0}^3\ell_i+\ell_i'+\ell_i''+\ell_*+\ell_1'+\ell_*''} \calD_{\ell_*\ell_*'\ell*''} \\
&&\,\times\, (-1)^{m_*+m_1'+m_*''} \left[\prod_{i=1}^3\calC^{\ell_{Gi} \ell_i' \ell_i''}_{m_{Gi} m_i' m_i''} \right] \calC^{\ell_1' \ell_2' \ell_3'}_{m_1' m_2' m_3'} \nonumber\\
&&\,\times\, \calC^{\ell_{G0} \ell_{G1} \ell_*}_{m_{G0} m_{G1} m_*} \calC^{\ell_* \ell_{G2} \ell_{G3} }_{-m_* m_{G2} m_{G3}}  \calC^{\ell_{0}'' \ell_{1}' \ell_*'}_{m_{0}'' m_{1}' m_*'} \calC^{\ell_*' \ell_{2}'' \ell_{3}'' }_{-m_*' m_{2}'' m_{3}''}. \nonumber
\end{eqnarray}
Depending on the position of the zero among $\ell_{Gi}$, for $i=\{0,1,2,3\}$, different solutions can in principle arise. 
We consider the following cases: (1) $\ell_{G0}=0$, where an entire column of the 9-j symbol vanishes; (2) $\ell_{G1}=0$, where two elements of the first 9-$j$ symbol are zero; and $\ell_{G2}=0$, where one element in each 9-$j$ symbol is zero. The fourth case, where $\ell_{G3}=0$, is trivially degenerate with case (3) and is therefore not treated separately.

For case (1), since we set $\ell_{G0}=0$, $\ell_0''\rightarrow 0$. Therefore, we have: 
\begin{eqnarray}\label{eqn:calQ_I}
\calQ^{\calL_G \Lambda^{'} \Lambda^{''}}_{\rm I} &=& (-1)^{\sum_{i=1}^3 \ell_{Gi}+\ell_i'+\ell_i''}  \sum_{M M' M''}\left[\prod_{i=1}^3\calC^{\ell_{Gi} \ell_i' \ell_i''}_{m_{Gi} m_i' m_i''} \right] \nonumber\\
&&\,\times\, \calC^{\ell_{G1} \ell_{G2} \ell_{G3}}_{m_{G1} m_{G2} m_{G3}} \calC^{\ell_{1}' \ell_{2}' \ell_{3}'}_{m_{1}' m_{2}' m_{3}'} \calC^{\ell_{1}'' \ell_{2}'' \ell_{3}''}_{m_{1}'' m_{2}'' m_{3}''}\nonumber\\
&=& \nine{\ell_{G1}}{\ell_{G2}}{\ell_{G3}}{\ell_1'}{\ell_2'}{\ell_3'}{\ell_1''}{\ell_2''}{\ell_3''},
\end{eqnarray}
where we relabel $\ell_*\rightarrow\ell_{G1}$, $\ell_*'\rightarrow\ell_{1}'$, and $\ell_*''\rightarrow\ell_{1}''$. The phase ${(-1)^{\ell_{G1}+\ell_1'+\ell_1''}}$ vanishes due to the presence of $\calC^{\ell_{G1} \ell_1' \ell_1''}_{000}$.
Given the relabelling, we find
\begin{eqnarray}\label{eqn:calD_calC_Lambda_I}
\calD_{\Lambda''}\calC^{\Lambda''}_{0} &=& \calD_{\ell_1'' \ell_2'' \ell_3''} \sqrt{2\ell_1''+1} (-1)^{\ell_1''} \calC^{0\ell_1'' \ell_1''}_{000} \calC^{\ell_1'' \ell_2'' \ell_3''}_{000} \nonumber\\
&=& \calD_{\ell_1'' \ell_2'' \ell_3''} \calC^{\ell_1'' \ell_2'' \ell_3''}_{000}.
\end{eqnarray}
Combining Eqs.~(\ref{eqn:calQ_I} -- \ref{eqn:calD_calC_Lambda_I}), the prefactor for case (1) reads
\begin{eqnarray}\label{eqn:calF_Lambda_I}
\calF^{\calL_G \Lambda^{'} \Lambda^{''}}_{\rm I} 
&=& (4\pi)^{-2} \calE_G \prod_{i=1}^3 \calD_{\ell_{Gi}\ell_i'\ell_i''} \calC^{\ell_{Gi}\ell_i'\ell_i''}_{000}\\
&&\,\times\,\nine{\ell_{G1}}{\ell_{G2}}{\ell_{G3}}{\ell_1'}{\ell_2'}{\ell_3'}{\ell_1''}{\ell_2''}{\ell_3''} \calD_{\ell_1'' \ell_2'' \ell_3''} \calC^{\ell_1'' \ell_2'' \ell_3''}_{000}\nonumber
\end{eqnarray}
For case (2), we set $\ell_{G1}=0$, therefore $\ell_{G0}\rightarrow \ell_{G1}$, $\ell_0''\rightarrow \ell_{G1}$, $\ell_*'\rightarrow \ell_1'$, $\ell_1''\rightarrow \ell_1'$, and $\ell_*\rightarrow \ell_{G1}$. We find:
\begin{eqnarray}\label{eqn:calQ_II}
\calQ^{\calL_G \Lambda^{'} \Lambda^{''}}_{\rm II} &=& \sum_{M M' M''} (\calD_{\ell_{G1} \ell_1'})^{-2} \calD_{\ell_{G1}\ell_1'\ell_1''} \calC^{\ell_{G1} \ell_{G2} \ell_{G3}}_{m_{G1} m_{G2} m_{G3}} \calC^{\ell_{1}' \ell_{2}' \ell_{3}'}_{m_{1}' m_{2}' m_{3}'} \nonumber\\
&&\,\times\, \calC^{\ell_{G2} \ell_2' \ell_2''}_{m_{G2} m_2' m_2''} \calC^{\ell_{G3} \ell_3' \ell_3''}_{m_{G3} m_3' m_3''} \sum_{\ell_1'' m_1''} \calC^{\ell_{G1} \ell_1' \ell_1''}_{m_{G1} m_1' m_1''} \calC^{\ell_1'' \ell_2'' \ell_3''}_{-m_1'' m_2'' m_3''} \nonumber\\
&=& (\calD_{\ell_{G1} \ell_1'})^{-2} \calD_{\ell_{G1}\ell_1'\ell_1''} \nine{\ell_{G1}}{\ell_{G2}}{\ell_{G3}}{\ell_1'}{\ell_2'}{\ell_3'}{\ell_1''}{\ell_2''}{\ell_3''}
\end{eqnarray}
where we have relabelled $\{\ell_{0}'', \ell_*\}\rightarrow\ell_{G1}$ and $\{\ell_1'',\ell_*'\}\rightarrow\ell_1'$ in the first line; and $\ell_*'' \rightarrow \ell_1''$ for the last line, which leads to a prefactor of $(\calD_{\ell_{G1}\ell_1'})^2$. For case (2), we find
\begin{eqnarray}\label{eqn:calD_calC_Lambda_II}
\calD_{\Lambda''}\calC^{\Lambda''}_{0} &=& \calD_{\ell_1'' \ell_2'' \ell_3''} \calD_{\ell_{G1} \ell_1'} (-1)^{\ell_1''} \calC^{\ell_{G1} \ell_1' \ell_1''}_{000} \calC^{\ell_1'' \ell_2'' \ell_3''}_{000}.
\end{eqnarray}
Combining Eqs.~(\ref{eqn:calQ_II} -- \ref{eqn:calD_calC_Lambda_II}), the prefactor for case (2) becomes
\begin{eqnarray}
&&\calF^{\calL_G \Lambda^{'} \Lambda^{''}}_{\rm II}\nonumber\\
&=& (4\pi)^{-2} \calE_G\,
\calD_{\ell_{G1} \ell_1' \ell_1''} \calD_{\ell_{G1} 0 \ell_{G1}} \calC^{\ell_{G1} 0 \ell_{G1}}_{000}
\calD_{0 \ell_{1}' \ell_{1}'} \calC^{0 \ell_{1}' \ell_{1}'}_{000}\nonumber\\
&&\times \prod_{i=2}^3 \calD_{\ell_{Gi}\ell_i'\ell_i''} \calC^{\ell_{Gi}\ell_i'\ell_i''}_{000} \left(\calD_{\ell_{G1}\ell_1'}\right)^{-2} \nine{\ell_{G1}}{\ell_{G2}}{\ell_{G3}}{\ell_1'}{\ell_2'}{\ell_3'}{\ell_1''}{\ell_2''}{\ell_3''} \nonumber\\
&&\,\times\, \calD_{\ell_1'' \ell_2'' \ell_3''} \calD_{\ell_{G1} \ell_1'} (-1)^{\ell_1''} \calC^{\ell_{G1} \ell_1' \ell_1''}_{000} \calC^{\ell_1'' \ell_2'' \ell_3''}_{000},
\end{eqnarray}
since $\calC^{\ell_{G1} 0 \ell_{G1}}_{000} = (-1)^{\ell_{G1}}\calD^{-1}_{\ell_{G1}}$ and $\calC^{\ell_{1}' 0 \ell_{1}'}_{000} = (-1)^{\ell_{1}'}\calD^{-1}_{\ell_{1}'}$, the coefficients cancels with $\calD_{\ell_{G1}\ell_1'}$ which arises in Eq.~\eqref{eqn:calD_calC_Lambda_II}.
The resulting phase $(-1)^{\ell_{G1}+\ell_1'+\ell_1''}$ also cancels due to the presence of $\calC^{\ell_{G1}\ell_1'\ell_1''}_{000}$. As a result, we have 
\begin{eqnarray}\label{eqn:calF_Lambda_II}
\calF^{\calL_G \Lambda^{'} \Lambda^{''}}_{\rm II}
&=& (4\pi)^{-2} \calE_G \prod_{i=1}^3 \calD_{\ell_{Gi}\ell_i'\ell_i''} \calC^{\ell_{Gi}\ell_i'\ell_i''}_{000} \\
&&\,\times\, \nine{\ell_{G1}}{\ell_{G2}}{\ell_{G3}}{\ell_1'}{\ell_2'}{\ell_3'}{\ell_1''}{\ell_2''}{\ell_3''}\calD_{\ell_1'' \ell_2'' \ell_3''} \calC^{\ell_1'' \ell_2'' \ell_3''}_{000}\nonumber,    
\end{eqnarray}
where case (1) and (2) share the same coefficients $\calF^{\calL_G \Lambda^{'} \Lambda^{''}}$.

\begin{figure}
    \centering
    \includegraphics[width=0.7\linewidth]{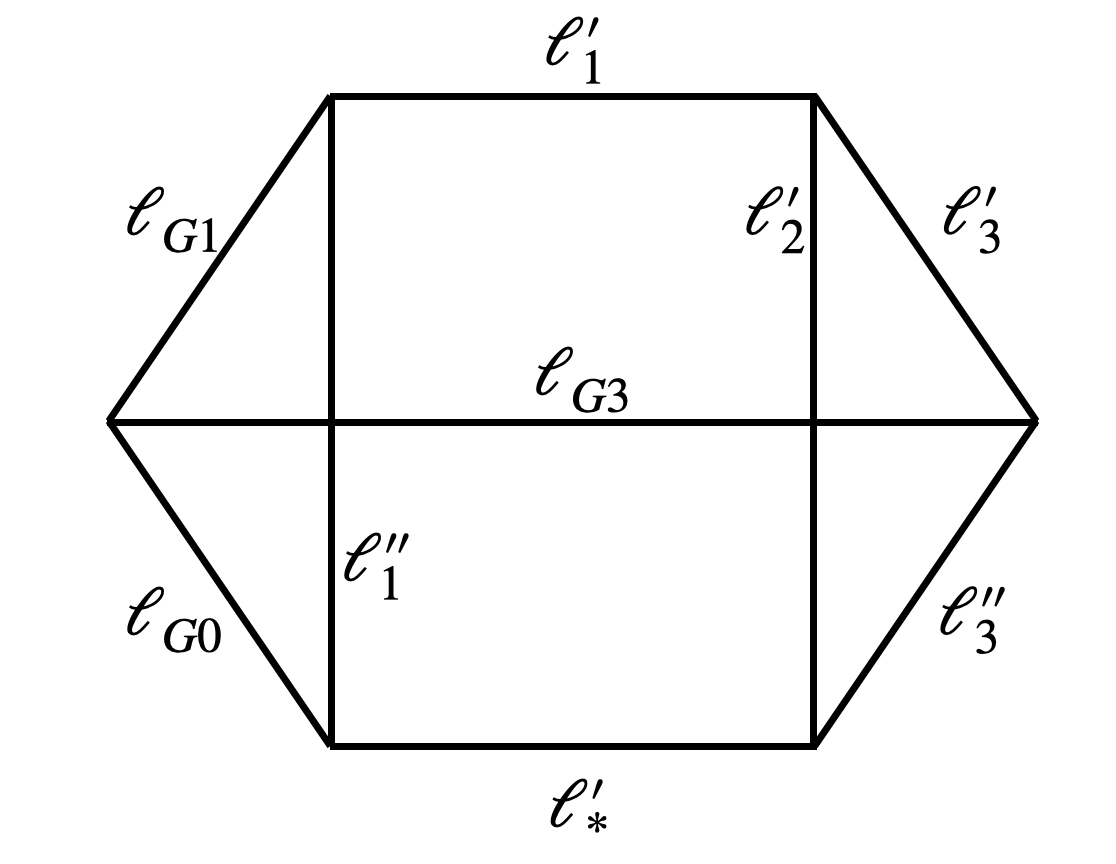}
    \caption{Yutsis diagram for the type-II 9-$j$ symbol.}
    \label{fig:9j-typeII}
\end{figure}

For case (3), we set $\ell_{G2}=0$, such that $\ell_{G0}=\ell_{G2}$, $\ell_0''=\ell_{G2}$, $\ell_*'=\ell_1'$, $\ell_*=\ell_{G3}$, $\ell_2''=\ell_2'$. We therefore find:
\begin{eqnarray}\label{eqn:calQ_III}
\calQ^{\calL_G \Lambda^{'} \Lambda^{''}}_{\rm III}
&=& \calD^{-1}_{\ell_{G2}\ell_1'\ell_2'\ell_{G3}} \calD_{\ell_{G3}\ell_1'\ell_2''}(-1)^{\ell_{G2}+\ell_1'+\ell_2'+\ell_{G3}} \nonumber\\
&&\,\times\!\sum_{\substack{\ell_*\ell_*'', m_* m_*''}}(-1)^{\sum_{i=0}^3\ell_i+\ell_i'+\ell_i''+\ell_{G3}+\ell_1'+\ell_2''}\nonumber\\
&&\,\times\,\sum_{m's} (-1)^{m_{G2}+m_1'+m_2'+m_{G3}} (-1)^{m_{G3}+m_1'+m_2''} \,  \nonumber\\
&& \,\times\, \calC^{\ell_{G1}\ell_1' \ell_1''}_{m_{G1}m_1' m_1''}\calC^{\ell_{G2}\ell_{G1}\ell_{G3}}_{m_{G2}m_{G1}m_{G3}}  \calC^{\ell_1' \ell_2' \ell_3'}_{m_1' m_2' m_3'} \calC^{\ell_{G2}\ell_1'' \ell_*''}_{m_{G2}m_1'' m_*''} \nonumber\\
&&\,\times\,\calC^{\ell_*'' \ell_2' \ell_3''}_{-m_*'' m_2' m_3''} \calC^{\ell_{G3}\ell_3' \ell_3''}_{m_{G3}m_3' m_3''},   
\end{eqnarray}
where the 3-$j$ connection structure is represented by the type-II Yutsis diagram in Fig.~\ref{fig:9j-typeII}. Here all the phases involving summation of angular momenta $\ell$'s and $m$'s all cancel with each other.

To restore the 9-$j$ symbol, we need to rearrange the product of the 3-$j$ symbols: $\calC^{\ell_{G2}\ell_1''\ell_*''}_{m_{G2}m_1''m_*''}\calC^{\ell_{*}''\ell_2'\ell_3''}_{-m_{*}''m_2'm_3''}$. In order to do so, we first recognize the property of the integral for the product of four spherical harmonics:
\begin{eqnarray}\label{eqn:prod_4_Ylm}
&&\int d\hn\, Y_{\ell_{G2}m_{G2}}(\hn) Y_{\ell_{1}''m_{1}''}(\hn) Y_{\ell_{2}'m_{2}'}(\hn) Y_{\ell_{3}''m_{3}''}(\hn)\nonumber \\
&=& \sum_{\ell_*'' m_*''} (-1)^{m_*''} \calG^{\ell_{G2}\ell_2'\ell_*''}_{m_{G2}m_2'm_*''}\calG^{\ell_{*}''\ell_1'\ell_3''}_{-m_{*}''m_1'm_3''}\nonumber\\
&=& \sum_{\ell_*'' m_*''} (-1)^{m_*''} \calG^{\ell_{G2}\ell_1'\ell_*''}_{m_{G2}m_1'm_*''}\calG^{\ell_{*}''\ell_2'\ell_3''}_{-m_{*}''m_2'm_3''},
\end{eqnarray}
where the Gaunt integral $\calG^{\ell_1\ell_2\ell_3}_{m_1 m_2 m_3}$ involves product of 3-$j$ symbols both with all $m$'s being zero $m_1=m_2=m_3=0$ as well as the ones being non-zero. For the 3-$j$ symbols with all $m$'s being zero, they are defined as 
\begin{eqnarray}
\left(\begin{array}{ccc}
\ell_1 & \ell_2 & \ell_3 \\
0 & 0 & 0 
\end{array}\right)
= \begin{cases}0, & J \text { odd }, \\
 \frac{(-1)^{L/2} (L/2)!}{\left((L+1)!\right)^{1/2}}\frac{\left(\left(L-2 \ell_i\right)!\right)^{1/2}}{(L/2-\ell_i)!} , & J \text { even. }\end{cases}\nonumber\\
\end{eqnarray}
where $L\equiv \ell_1+\ell_2+\ell_3$. As a result, we can re-order the arrangement of the product 
\begin{eqnarray}
\calC^{\ell_{G2}\ell_2'\ell_*''}_{000}\calC^{\ell_*''\ell_2'\ell_3''}_{000} = \calC^{\ell_{G2}\ell_2'\ell_*''}_{000}\calC^{\ell_*''\ell_2'\ell_3''}_{000}
\end{eqnarray}
Combining with Eq.~\eqref{eqn:prod_4_Ylm}, we find that
\begin{eqnarray}
\calC^{\ell_{G2}\ell_2'\ell_*''}_{m_{G2}m_2'm_*''}\calC^{\ell_{*}''\ell_1'\ell_3''}_{-m_{*}''m_1'm_3''} = \calC^{\ell_{G2}\ell_1'\ell_*''}_{m_{G2}m_1'm_*''}\calC^{\ell_{*}''\ell_2'\ell_3''}_{-m_{*}''m_2'm_3''}.
\end{eqnarray}
By relabeling $\ell_*'' \rightarrow \ell_2''$, case-III can be written as
\begin{eqnarray}\label{eqn:calQ_III_v2}
\calQ^{\calL_G \Lambda^{'} \Lambda^{''}}_{\rm III} 
&=& \calD^{-1}_{\ell_{G2}\ell_2'} \calD_{\ell_2''} \nine{\ell_{G1}}{\ell_{G2}}{\ell_{G3}}{\ell_1'}{\ell_2'}{\ell_3'}{\ell_1''}{\ell_2''}{\ell_3''} \nonumber\\
\end{eqnarray}
For case (3), we find
\begin{eqnarray}\label{eqn:calD_calC_Lambda_III}
\calD_{\Lambda''}\calC^{\Lambda''}_{0} \!&=&\! \calD_{\ell_1'' \ell_2'' \ell_3''} \calD_{\ell_{G2} \ell_2'} (-1)^{\ell_2''} \calC^{\ell_{G2} \ell_2' \ell_1''}_{000} \calC^{\ell_1'' \ell_2'' \ell_3''}_{000}.
\end{eqnarray}
Combining Eqs.~(\ref{eqn:calQ_III_v2} -- \ref{eqn:calD_calC_Lambda_III}), we find
\begin{eqnarray}\label{eqn:calF_Lambda_III}
\calF^{\calL_G \Lambda^{'} \Lambda^{''}}_{\rm III}
&=& (4\pi)^{-2} \calE_G \prod_{i=1}^3 \calD_{\ell_{Gi}\ell_i'\ell_i''} \calC^{\ell_{Gi}\ell_i'\ell_i''}_{000} \\
&&\,\times\, \nine{\ell_{G1}}{\ell_{G2}}{\ell_{G3}}{\ell_1'}{\ell_2'}{\ell_3'}{\ell_1''}{\ell_2''}{\ell_3''}\calD_{\ell_1'' \ell_2'' \ell_3''} \calC^{\ell_1'' \ell_2'' \ell_3''}_{000}\nonumber,    
\end{eqnarray}
Comparing Eq.~\eqref{eqn:calF_Lambda_I}, Eq.~\eqref{eqn:calF_Lambda_II}, and Eq.~\eqref{eqn:calF_Lambda_III}, we find that the coefficients can be unified, regardless of where the zero appears in the 3n-$j$ symbols.

\bibliographystyle{aipnum4-2}
\bibliography{ref.bib} 

\end{document}